\documentclass{aa}
\usepackage{hyperref}
\usepackage{txfonts}
\usepackage{graphicx}
\usepackage{amsmath}
\usepackage{tabularx}
\usepackage{enumitem}
 
\usepackage{natbib}
\bibpunct{(}{)}{;}{a}{}{,}
\usepackage{color}

\begin{document}

\title{Broadband study of blazar 1ES 1959+650 during\\ 
flaring state in 2016}

\author{S. R. Patel\inst{1}, A. Shukla\inst{2}, V. R. Chitnis\inst{1}, 
D. Dorner\inst{2}, K. Mannheim\inst{2}, B. S. Acharya\inst{1}, 
B. J. Nagare\inst{3}} 

\institute{Department of High Energy Physics, Tata Institute of Fundamental 
Research, Mumbai 400005, India
     \and Universit$\ddot{a}$t W$\ddot{u}$rzburg, 97074 W$\ddot{u}$rzburg, 
Germany 
     \and Department of Physics, University of Mumbai, Santacruz (East), 
Mumbai-400098, India}
\date{}

\abstract
{}
{The nearby TeV blazar 1ES 1959+650 (z=0.047) was reported to be in flaring state 
during June - July 2016 by Fermi-LAT, FACT, MAGIC and VERITAS collaborations. 
We studied the spectral energy distributions (SEDs) in different states of 
the flare during MJD 57530 - 57589 using simultaneous multiwaveband data to 
understand the possible broadband emission scenario during the flare.}
{The UV/optical and X-ray data from UVOT and XRT respectively on board Swift and high energy $\gamma$-ray 
data from Fermi-LAT are used to generate multiwaveband lightcurves as well as
to obtain high flux states and quiescent state SEDs. 
The correlation and lag between different energy bands is quantified using 
discrete correlation function. The synchrotron self 
Compton (SSC) model was used to reproduce the observed SEDs during 
flaring and quiescent states of the source.}
{A decent correlation is seen between X-ray and high energy $\gamma$-ray fluxes. The spectral hardening 
with increase in the flux is seen in X-ray band. The powerlaw index vs flux plot in
$\gamma$-ray band indicates the different emission regions for 0.1 - 3 GeV and 3-300 GeV energy photons.
Two zone SSC model satisfactorily fits the observed broadband 
SEDs. The inner zone is mainly responsible for producing synchrotron peak and high 
energy $\gamma$-ray part of the SED in all states. The second zone is mainly required to 
produce less variable optical/UV and low energy $\gamma$-ray emission.}
{Conventional single zone SSC model does not
satisfactorily explain broadband emission during observation period considered. There is an 
indication of two emission zones in the jet which are responsible for producing broadband
 emission from optical to high energy $\gamma$-rays.}

\keywords{Radiation mechanisms: non-thermal --
          Galaxies: BL Lacerate objects: individual: 1ES 1959+650 --
          Gamma rays: general --
          X-rays: galaxies
          }

\titlerunning{Spectral and temporal studies of 1ES 1959+650}
\authorrunning{S. R. Patel et al.}

\maketitle

\section{Introduction}
  
Blazars are a subclass of Active Galactic Nuclei (AGN) having relativistic 
jets pointing close to our line of sight \citep{urry}. Jets emit highly variable 
non-thermal radiation spanning wide band of frequencies from radio to $\gamma$-rays. 
Blazars include two types of objects, BL Lacreate (BL Lac) characterized by 
featureless optical spectra and flat spectrum radio quasars (FSRQ) which 
show prominent emission lines.

The spectral energy distributions (SEDs) of these objects show characteristic 
two hump structure. The first low frequency hump is attributed to the synchrotron radiation from relativistic electrons, while 
second high frequency hump is understood as possibly corresponding to inverse compton scattering of these synchrotron 
photons (SSC i.e. Synchrotron Self-Compton) or external photons (EC i.e. External Compton) by the same population of electrons. 
Alternative explanation for origin of the second hump is given in terms of
hadronic models including neutral pion decay, proton synchrotron 
\citep{karl1998} etc. Comprehensive review of these mechanisms 
is given by \citet{Bottcher}.

Depending on the position of synchrotron peak in SED, BL Lacs are further 
divided into three classes \citep{pado}. Low frequency BL Lac (LBL) objects 
exhibit synchrotron peak in IR-Optical band, intermediate frequency BL Lac 
objects (IBL) have their synchrotron peak at optical-UV frequencies and 
high frequency BL Lac (HBL) objects show synchrotron peak in UV - X-ray band. 

The HBL object, 1ES 1959+650 (z = 0.047) was first detected in radio band 
with NARO Green Bank 91 m telescope \citep{radio, radio2} and later 
in X-ray \citep{5X-ray} using Imaging Proportional counter (IPC) on board 
Einstein Observatory. TeV emission from this source was first observed by 
Utah Seven Telescope Array collaboration with total significance of 3.9$\sigma$ 
above 600 GeV \citep{nish}. The source was later observed during 2002 May 16 to July 8,
 with strong detection significance of > 20$\sigma$ by Whipple 10 m telescope \citep{hold}.  
Since then it has shown several flaring episodes at VHE ( > 100 GeV) $\gamma$-ray 
energies, with the most noticeable one being in 2002 when it showed enhanced 
TeV emission without any contemporaneous X-ray flare \citep{orpahan,orpkra,orpdan,orprei}. 
In 2004, the source was observed in low state by MAGIC collaboration with a flux of about 20\% of the Crab and 
at $\sim$  8$\sigma$ significance  level above $\sim$ 180 GeV \citep{6magic}. 
Broadband variability of the source was studied in 2012 using strictly 
simultaneous observations from VERITAS and Swift and reflected emission 
scenario was used to explain the variability \citep{aliu2014}. The 
preliminary analysis of data from Fermi-LAT and various ground based Cherenkov
experiments such as FACT, MAGIC, VERITAS as reported by \citet{atel}, indicated 
flaring activity in the source 27 April 2016 onwards.

In the present paper, we have examined the multiwaveband emission from this 
source over 800 days during MJD 57000 to 57800 (9 December 2014 to 16 February
 2017) and studied the broadband variability of 
this source for the period from MJD 57530 to 57589 (22 May 2016 - 20 July 2016)
during which it showed increased flux in X-ray, Fermi-LAT (100 MeV - 300 GeV) 
and TeV bands. To have good statistics in LAT energy band (0.1 -300 GeV), we chose six 
periods of 10 days each to sample the complete 
flare and investigated its emission mechanism in different states during 
the flare using SSC model. We also studied SED 
corresponding to 10 days period when source was in low state.
The paper is organized as follows. In section 2 various data sets and
analysis methods are described. In section 3 timing and spectral studies 
are elaborated. The SED modeling is outlined in section 4 followed 
by discussion and conclusions in section 5.

\section{Multiwaveband  observations and analysis}

We have studied the multiwaveband data from radio to $\gamma-$rays spanning 
the period of more than two years from 9 December 2014 (MJD 57000) to 16 
February 2017 (MJD 57800). We analyzed UV-optical data from Swift-UVOT, X-ray 
data from Swift-XRT and high energy $\gamma-$ray data from Fermi-LAT.  We 
also used publicly available data from OVRO, SPOL, MAXI and Swift-BAT. Details 
of these data sets and analysis procedure are given below.

\begin{figure*}
  \includegraphics[angle=0,width=20cm,height=23cm]{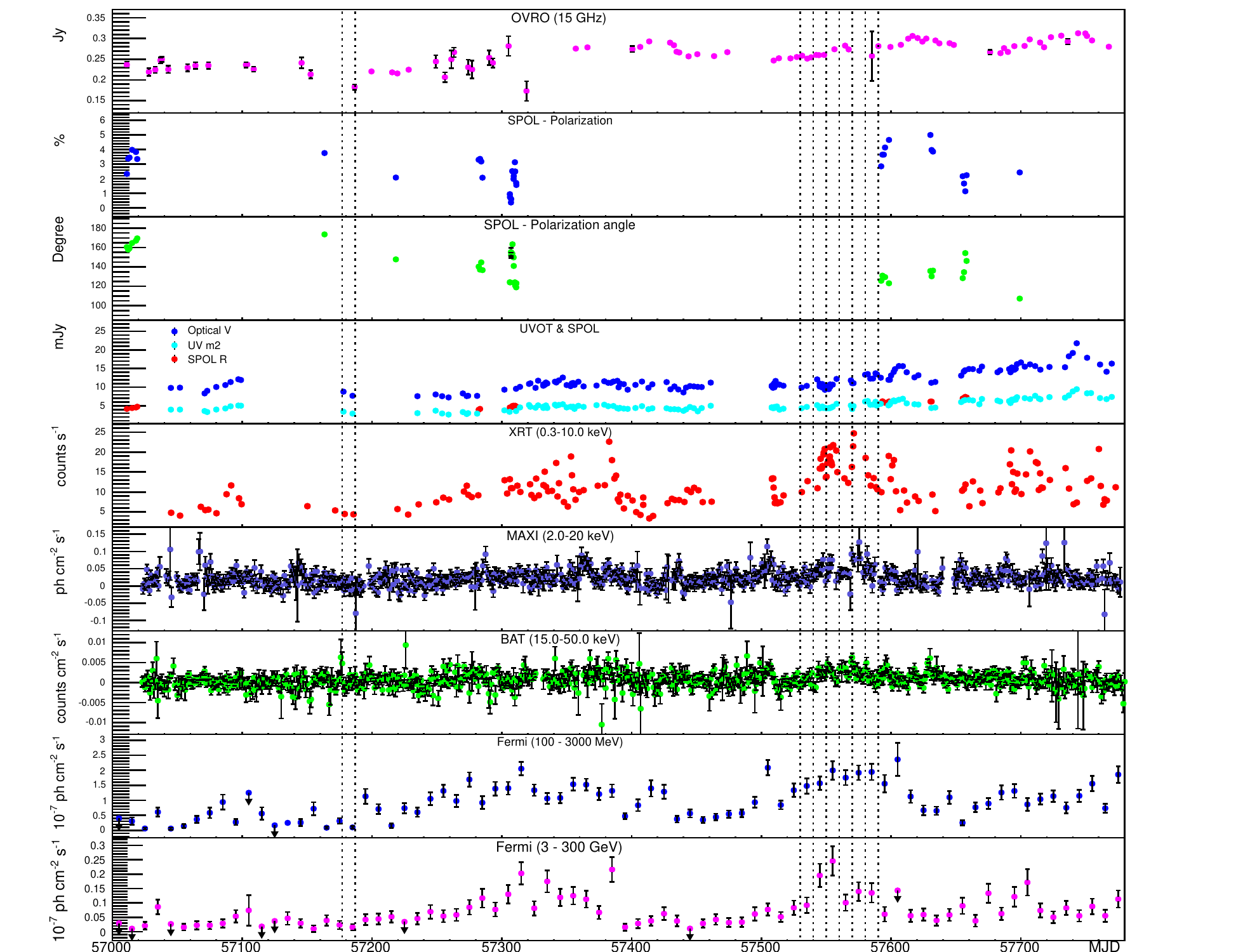}
   \caption{Two years light curves of 1ES 1959+650 from MJD 57000 to 57800, 
Panel-1 : Radio flux density at 15 GHz in Jy  ; Panel-2 : Degree of polarization 
from CCD-SPOL observations , Panel-3 : Polarization angle from CCD-SPOL
observations, Panel-4 : UVOT-U, UVOT-W2 band and CCD-SPOL-R band fluxes in mJy 
(Corrected for host galaxy contribution), Panel-5 : Swift-XRT count rate in counts/s, 
Panel-6 : MAXI flux in photons $cm^{-2} s^{-1}$ (daily average), Panel-7 : 
Swift-BAT flux in counts $cm^{-2} s^{-1}$(daily average), Panel-8 : Fermi-LAT 
0.1-3 GeV flux in photons $cm^{-2} s^{-1}$ (10 days average), Panel-9 : 
Fermi-LAT 3-300 GeV flux in photons $cm^{-2} s^{-1}$ (10 days average) ; 
SEDs are computed for the periods marked by dotted lines.}
  
   \label{FigMWLC}   
\end{figure*}

\subsection{High energy $\gamma-$ray observations}

High energy $\gamma-$ray data covering the energy range of 100 MeV - 300 
GeV was obtained from Large Area Telescope (LAT) on board Fermi spacecraft
\citep{fermi}. Data were analyzed using Science Tools version v10r0p5. User 
contributed enrico package \citep{enrico} 
was used. The events were extracted from the circular region of interest (ROI) 
of $20^{\circ}$ centered on the source. Zenith angle cut of $90^{\circ}$ 
was applied to filter the background $\gamma$-rays from Earth's limb. 
To select good time intervals, filter with '(DATA\_QUAL>0)\&\&(LAT\_CONFIG==1)' 
was used.  The spectral analysis was carried out using isotropic emission 
model (iso\_P8R2\_SOURCE\_V6\_v06.txt) and galactic diffuse emission  
component model (gll\_iem\_v06.fit) with post launch instrument response 
function (P8R2\_SOURCE\_V6) and using unbinned likelihood analysis.
The sources lying within ROI of 15$^\circ$ radius around the 1ES 1959+650 
from the 3FGL catalog were included in the model XML file. In likelihood fit, 
both spectral and normalization parameters of the sources within 5$^\circ$ 
radius around the source were left free to vary while keeping parameters for 
all other sources fixed at their catalog value. There are 72 point sources and 
1 extended source in the model file. The parameters of extended source were 
also kept free in maximum likelihood analysis. Source spectrum was modeled with 
a power law. The lightcurves were generated with 10 days binning in energy range
of 0.1 -3 GeV and 3 - 300 GeV.

\subsection{X-ray observations}

Publicly available hard X-ray data from Burst Alert Telescope (BAT) 
on board Swift are used \footnote{BAT : http://swift.gsfc.nasa.gov/results/transients/weak/1ES1959p650.lc.txt}. 
These are daily average count rates over the energy 
range of 15-50 keV. Soft X-ray data covering the energy range of 2-20 keV 
from Monitor of All sky X-ray Images (MAXI) on board International Space Station 
(ISS, \citet{iss}) are obtained from MAXI website 
\footnote{MAXI : http://134.160.243.77/star\_data/J1959+651/
J1959+651\_00055054g\_lc\_1day\_all.dat}. These are daily average flux values. 
 
Also soft X-ray data covering the energy range of 0.3 - 8 keV from X-ray 
telescope (XRT) on board Swift have been used \citep{xrt}. The data were 
analyzed using XRT data analysis software (XRTDAS) distributed within the 
HEASOFT package (v6.19). The xrtpipeline-0.13.2 tool was used to generate the 
cleaned event files. Data from 34 observations during 22 April - 20 
July 2016 corresponding to flare region were analyzed. The source and 
background data were extracted from the circular region of 20 pixels 
radius around the source and 40 pixels radius region away from the source 
respectively. The spectral data were combined into six groups of 10 days 
each as shown in Fig.~\ref{FigMWLC} and rebinned with minimum 20 photons 
per bin. Six spectra were fitted with absorbed powerlaw model as 
well as with log parabola model. Spectral form of log parabola model is 
given by

\begin{equation}
dN/dE = K(E/E_b)^{-\alpha-\beta log(E/E_b)}
\end{equation}

where $\alpha$ is the spectral index and $E_p$ is the point of maximum
curvature given by

\begin{equation}
E_p=E_b 10^{(2-\alpha)/2\beta}
\end{equation}

While fitting the spectrum, $E_b$ is fixed at 1 keV. To correct for interstellar
absorption of soft X-rays along line of sight, neutral hydrogen column density 
($N_H$) is fixed at 1.0 $\times 10^{21} cm^{-2}$ \citep{nh}. Swift XRT light curve 
spanning two years data, over the energy range of 0.3-10 keV is shown in 
Fig.~\ref{FigMWLC}.
This is publicly available Swift-XRT lightcurve obtained from 
the website\footnote{XRT :  http://www.swift.psu.edu/monitoring/data/
1ES1959+650/lightcurve2.txt}.

\subsection{UV, optical and radio observations}

We have analyzed Swift-UVOT \citep{uvot} data for the period of two years. 
Data are available in six different filters covering optical and UV band, 
viz. V, B, U, UVW1, UVM2 and UVW2. For each filter, images were added using 
tool uvotimsum and flux/magnitude values were obtained using tool uvotsource. 
For V, B and U filters, source counts were extracted from circular region 
with radius of 5" around the source location, whereas for UVW1, UVM2 and 
UVW2 filters, region with radius of 10" was used. The Galactic extinction 
correction \citep{corr} of $ E_{B-V} = 0.177$ mag was applied to observed 
magnitude. Observed magnitudes were then converted into flux using zero 
point magnitudes \citep{poole}. Host galaxy contribution \citep{HG} of 1.1 
mJy, 0.4 mJy and 0.1 mJy were subtracted for V, B and U filter respectively.
No correction is applied at UV frequencies as host galaxy contribution is
negligible at these frequencies. Fig.~\ref{FigMWLC} shows lightcurves from optical U and ultravilolet M2 filters. 
Other UVOT bands show similar trend, hence only two bands are shown to avoid cluttering.

1ES 1959+650 is being monitored with SPOL CCD Imaging/Spectropolarimeter at  
steward observatory at University of Arizona \citep{smith} regularly
as a part of the Fermi multiwavelength support programme. The publicly
available optical R-band and V-band photometric and linear polarization 
data were obtained from the SPOL 
website\footnote{http://james.as.arizona.edu/$\sim$psmith/Fermi/}. 
The R-band fluxes are corrected for host galaxy contribution which is 0.84 mJy \citep{HG-R}.
From available data the degree of linear polarization and polarization angle were found to
vary between 0.37$\%$ to 3.98$\%$ and 107$^\circ$ to 173$^\circ$ respectively. 

1ES 1959+650 is also observed regularly in radio as a part of Fermi monitoring 
programme by Owens Valley Radio Observatory (OVRO; \citet{ovro}). 
The publicly available data at 15 GHz were used in lightcurve from OVRO 
website\footnote{OVRO : http://www.astro.caltech.edu/ovroblazars/
data.php?page=data\_return}.

\section{Results}

In this section we present results from multiwaveband temporal and spectral
studies of 1ES 1959+650 collected over two years period from MJD 57000 to 57800.

\subsection{Multiwaveband temporal studies}

Fig.~\ref{FigMWLC} shows the lightcurve of 1ES 1959+650 for the period 
starting from MJD 57000 to MJD 57800. Panels corresponding to various 
wavebands have been arranged in increasing order of frequency from top 
to bottom, starting with radio lightcurve from OVRO in the topmost panel to high energy 
$\gamma-$rays from Fermi-LAT in the bottom-most panel. Fermi-LAT flux
values over the energy ranges 0.1-3 GeV and 3-300 GeV shown in last 
two panels, are averaged over 10 days bins. Data from Swift-BAT and 
MAXI are averaged over a day, whereas Swift-XRT and Swift-UVOT data 
points correspond to individual obervations with typical duration of 
about hours/minutes. SPOL and OVRO have an integration time of a few 
seconds. The source exhibited the flare in 2016 during 
MJD 57530 - 57589 which is clearly seen in the figure particularly in 
X-ray and $\gamma-$ray bands. We studied correlated variability in various 
wavebands. In this work we have concentrated on
detailed studies during flaring episode around MJD 57530 - 57589 (22 May -
20 July 2016). This episode is divided into six periods of ten days each.
Quiescent state data during MJD 57177 - 57186 (4 -13 June 2015) is
compared with this flare.

\subsubsection{Variability}

We have studied variability of lightcurves in various wavebands on various
time scales. Variability is estimated in terms of fractional variability 
amplitude, $ F_{var} $ parameter \citep{fvar, chitnis}. This parameter 
estimates variability intrinsic to the source and is given by
\begin{equation}
F_{var} = \sqrt{\frac{S^2 - {\sigma_{err}}^2}{\bar{x}^2}}
\end{equation}

where $S^{2}$ is the sample variance, ${\sigma_{err}}^2$  the mean square error and 
$\bar{x}$  the unweighted sample mean. The error on the $F_{var}$ is given by

\begin{equation}
\sigma_{F_{Var}} = \sqrt{\left(\sqrt{\frac{1}{2N}} \frac{{\sigma_{err}}^2}{\bar{x}^2 F_{var}}\right)^2  + \left(\frac{\sqrt{{\sigma_{err}}^2}}{N} \frac{1}{\bar{x}} \right)^2}
\end{equation}

\begin{figure}
  \centering
  \includegraphics[angle=0,width=8cm]{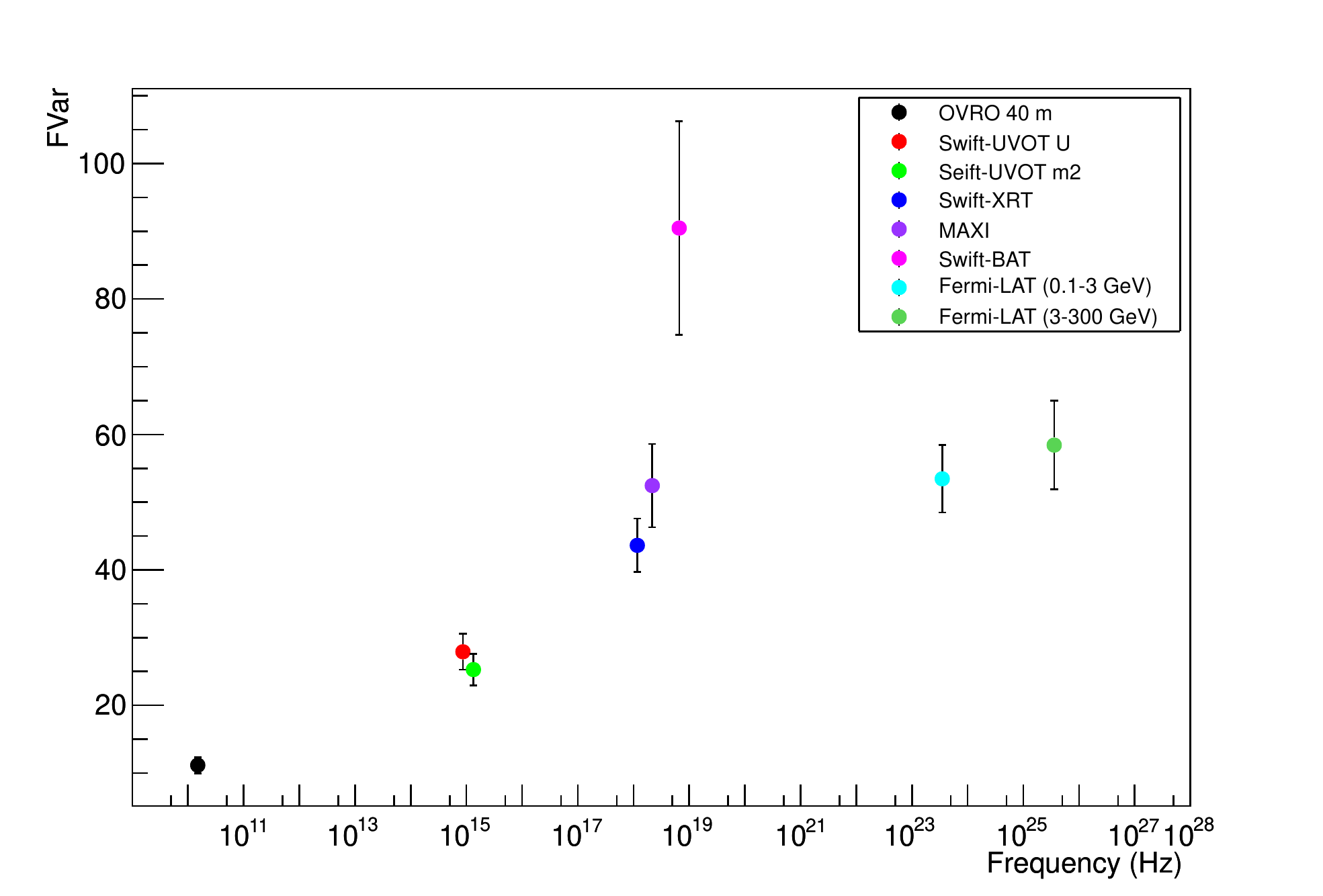}
  \caption{Fractional variability for 10 days binned lightcurve}
  \label{FigVar}
\end{figure}

Variability strength is estimated on time scales of 10 and 20 days in various
wavebands in present work. Results are given in Table. ~\ref{tab_var} and 
plotted in Fig.~\ref{FigVar} for 10 days binning.  The large error bar on BAT 
is due to poor sensitivity of the instrument. Variability seems to increase 
with frequency from radio to X-rays and decrease in high energy $\gamma-$rays
compared to hard X-rays. Similar trend was seen in Mkn 421 \citep{sinha16}.

\begin{table}
\caption{Fractional variability strength in various wavebands}

\begin{tabular}{ccc}
\hline
Waveband   & 10 days binning & 20 days binning \\
\hline

Radio (15 GHz)   & 11.16 +/- 1.19  & 11.46 +/- 1.59 \\
Optical U        & 27.91 +/- 2.65  & 26.96 +/- 3.34\\
UV M2            & 25.28 +/- 2.35  & 23.29 +/- 2.88\\
XRT (0.3 - 8 keV)& 43.62 +/- 3.95  & 38.27 +/- 4.58 \\
MAXI (2-20 keV)  & 52.45 +/- 6.13  & 48.30 +/- 6.84\\
BAT (15-50 keV)  & 90.47 +/- 15.78 & 96.39 +/- 16.66 \\
Fermi-LAT (0.1-3 GeV)  & 53.45 +/- 4.97 & 51.08 +/- 6.21 \\
Fermi-LAT (3-300GeV)   & 58.42 +/- 6.55 & 52.88 +/- 7.45 \\
\hline
\end{tabular}
\label{tab_var}
\end{table}

\subsubsection{Correlations}

The correlations between various lightcurves are quantified over the entire observation period (MJD 57000 to 57800) using discrete correlation 
function (DCF) \citep{edlson}. For two discrete data sets $a_{i}$ and $b_{j}$, the 
unbinned discrete correlation is defined as,
\begin{equation}
UDCF_{ij} = \frac{(a_{i} - \bar{a})(b_{j} - \bar{b})}{\sqrt{(\sigma_{a}^{2}-e_{a}^{2})(\sigma_{b}^{2}-e_{b}^{2})}} 
\end{equation}
for all measured pairs ($a_{i}$ ,$b_{j}$) having pairwise lag $\Delta t_{ij} = t_{j} - t_{i}$. 
$\sigma_{a}$ and $\sigma_{b}$ are standard deviations of each data train and 
$e_{a}$,$e_{b}$ are the measurement errors associated with them. DCF is then 
given by averaging M pairs for which 
$ (\tau - \Delta \tau/2) \leq \Delta t_{ij} < (\tau + \Delta \tau/2) $,
\begin{equation}
DCF(\tau) = \frac{1}{M}UDCF_{ij}
\end{equation} 

\begin{figure*}
\centering
  \includegraphics[angle=0,width=15cm,height=10cm]{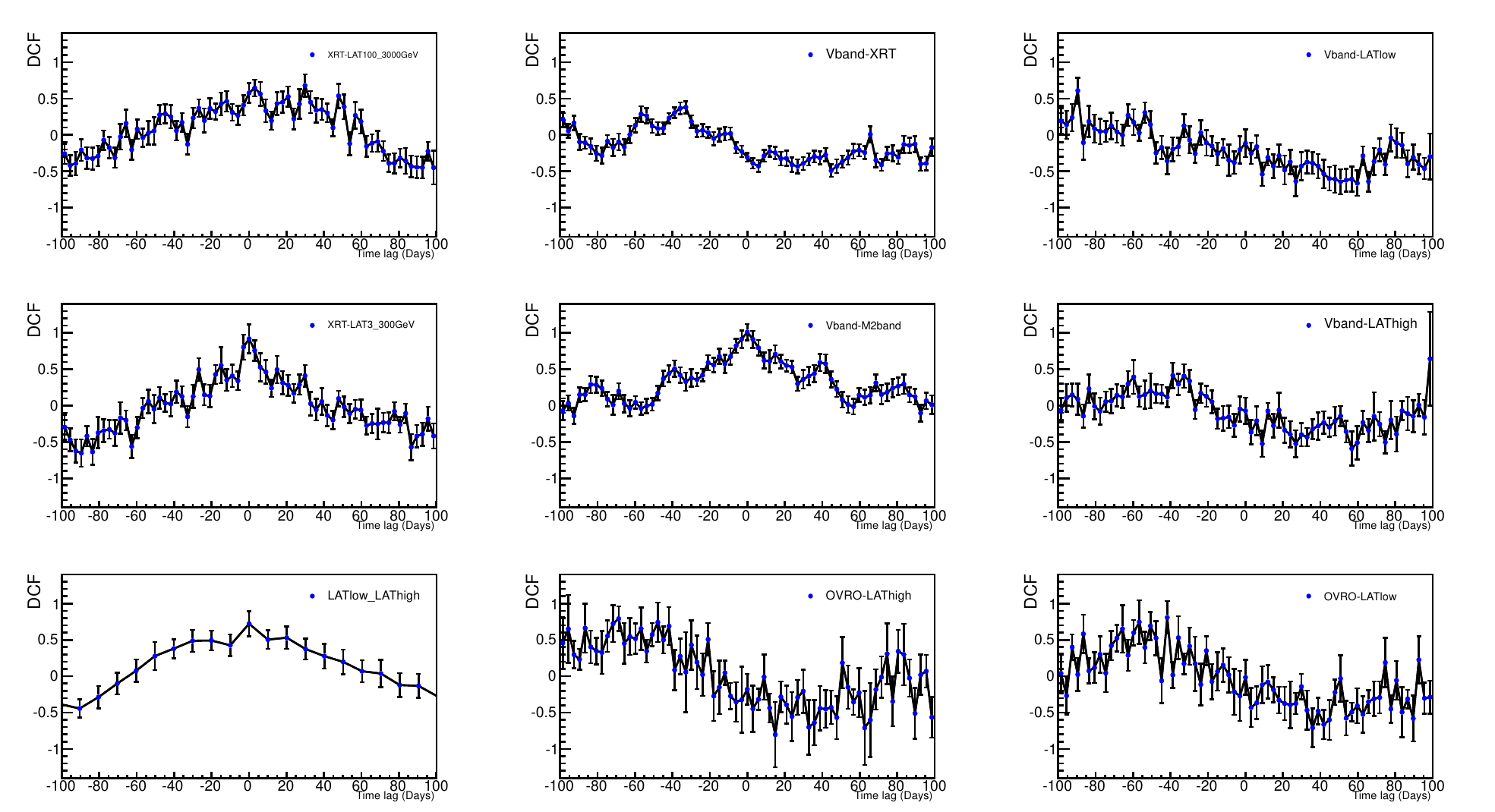}
  \caption{Discrete correlation function vs time lag between various energy bands}
  \label{FigZDCF_all}
\end{figure*}
The error on DCF is given by,
\begin{equation}
\sigma_{DCF}(\tau) = \frac{1}{M-1} \left\{ \sum_{}^{} [ UDCF_{ij} - DCF(\tau) ] \right\}^{1/2}
\end{equation}
Each data set is linearly detrended \citep{detrend} before using above function. 
The correlation coefficients along with lags between the lightcurves is listed in 
Table-~\ref{table_dcf} and their plots are shown in Fig. ~\ref{FigZDCF_all}.
Values of correlation coefficients given in table and figure are estimated for 3 days
binning in lag to improve statistics, except for 0.1-3 GeV -- 3-300 GeV case where bin
 size of 10 days is used. These values are consistent with one day binning of lag in all 
the cases. It can be seen that there is a strong correlation between optical (V band) 
and UV (M2 band) at zero lag. X-ray data shows good correlation
with low and high energy $\gamma$-rays with no visible lag. 
Correlation is better with higher energy $\gamma-$rays. Low and high energy $\gamma-$rays
are also correlated. The mild correlation is seen between optical
(V band) and $\gamma-$rays with a lag of about 38-90 days. Similar
lag is seen between radio and $\gamma-$rays. On the other hand, correlation
between optical and X-ray band is rather weak. This indicates X-ray and 
$\gamma-$rays, particularly higher energy ones, may have similar origin.
Whereas optical/UV band emission may have different origin. The study of 
long-term data (2005-2014) carried out by \citet{opt} showed that source 
displayed slow variation in optical R-band and exhibited optical flare 
lasting for several months to years. They also report relatively weak 
correlation between XRT and UVOT band. 
\begin{table}[!ht]
\caption{Maximum DCF between various wavebands}
\begin{tabular}{@{}ccc@{}}
\hline
Bands & DCF & Lag (Days) \\
\hline
XRT-0.1-3 GeV LAT  & 0.58 +/- 0.14 & 0.0 \\
XRT-3-300 GeV LAT  & 0.92 +/- 0.19 & 0.0 \\
0.1-3 GeV LAT-3-300 GeV LAT & 0.72 +/- 0.17 & 0.0 \\
UVOT (V)- XRT & 0.38 +/- 0.08 & -32.8 \\
OVRO (15 GHz)-0.1-3 GeV LAT & 0.81 +/- 0.22 & -41.8 \\
OVRO (15 GHz)-3-300 GeV LAT & 0.79 +/- 0.17 & -68.7 \\
UVOT (V)- UVOT (M2) & 1.00 +/- 0.11  & 0.0 \\
UVOT (V)-0.1-3 GeV LAT &  0.61 +/- 0.18 & -89.6 \\
UVOT (V)-3-300 GeV LAT & 0.41 +/- 0.17  & -38.8 \\
\hline
\end{tabular}
\label{table_dcf}
\end{table}
\subsubsection{Lognormality}
\begin{figure*}
   \centering
   \includegraphics[angle=0,width=6.5cm,height=4cm]{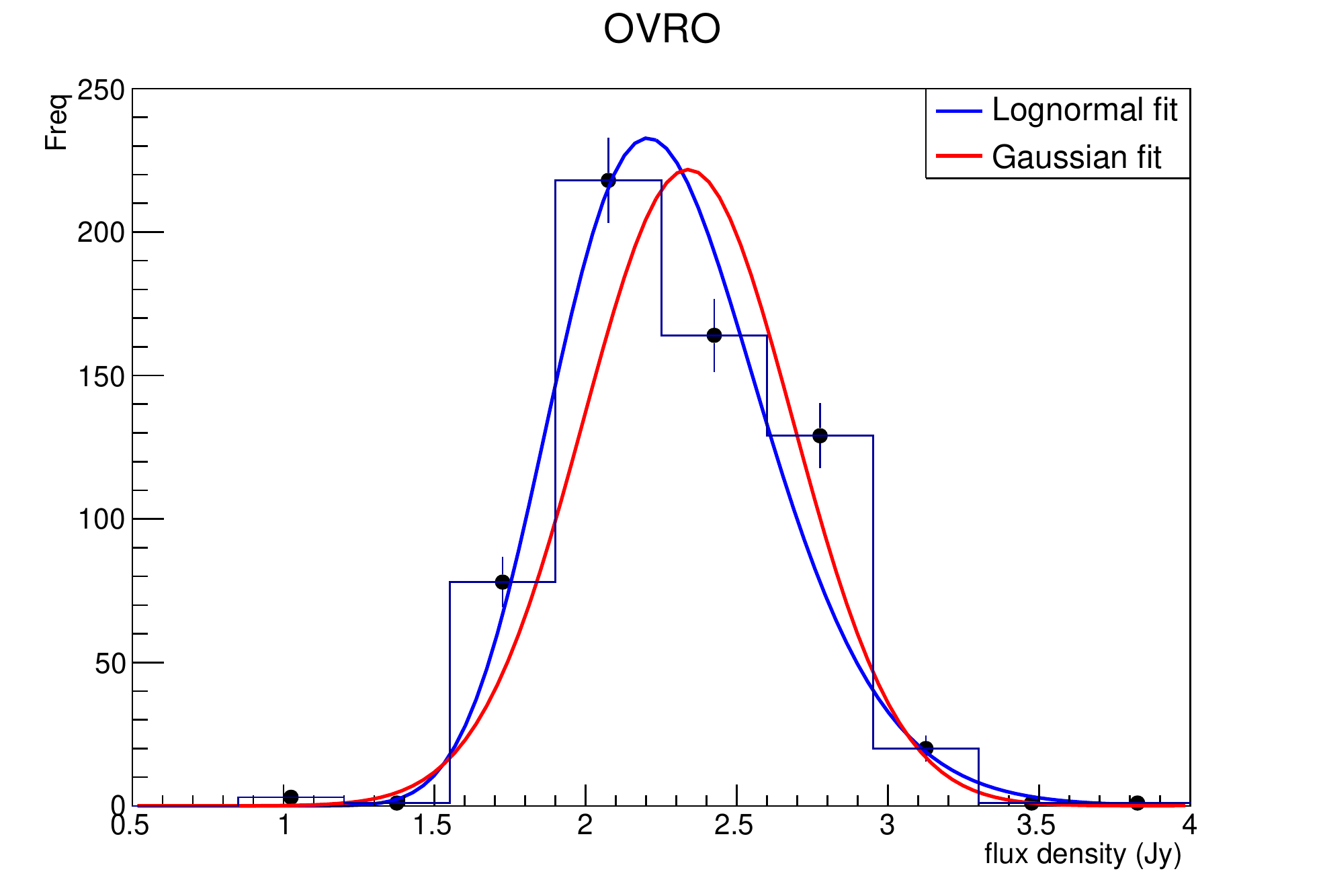} 
   \includegraphics[angle=0,width=6.5cm,height=4cm]{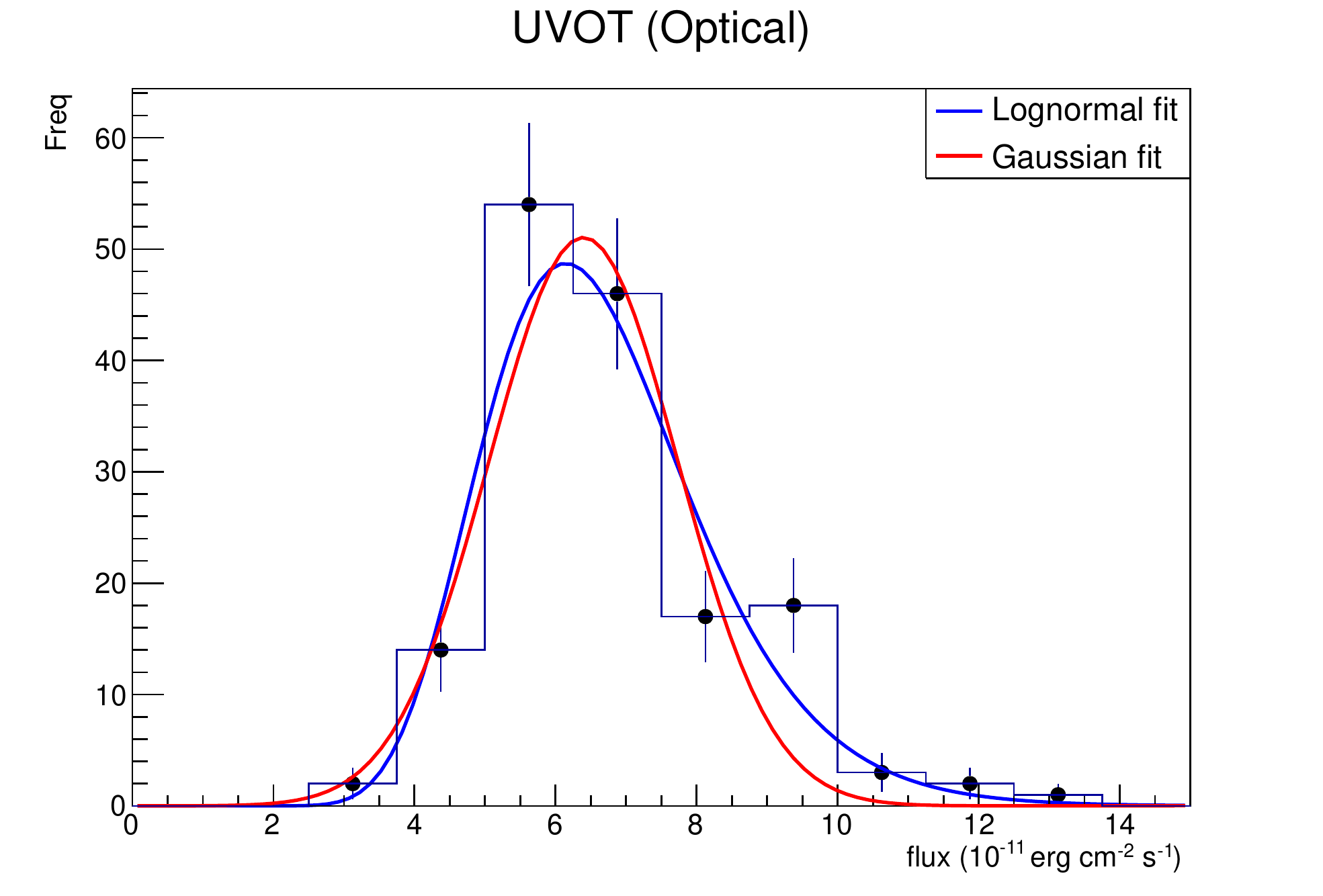}
   \includegraphics[angle=0,width=6.5cm,height=4cm]{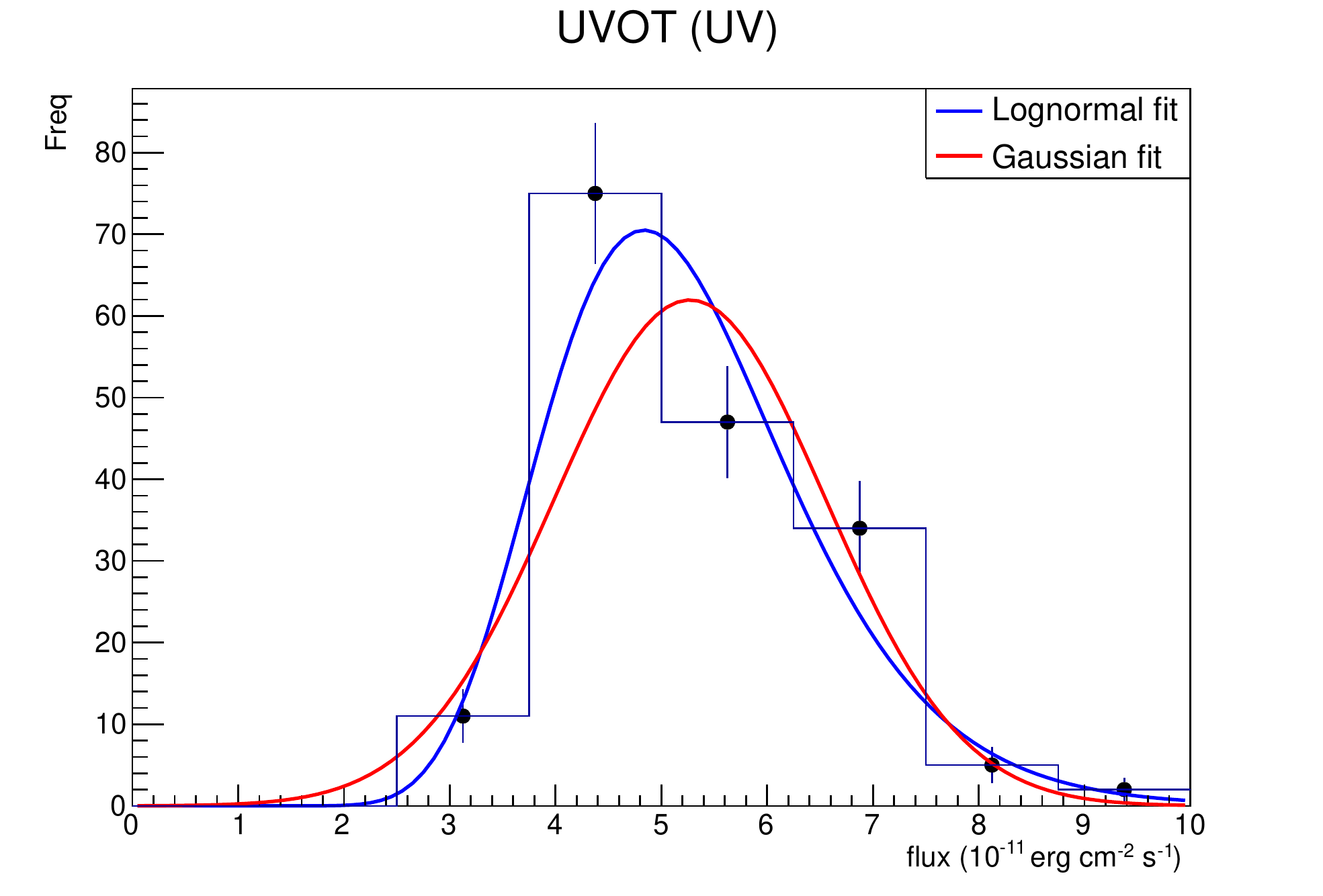}
   \includegraphics[angle=0,width=6.5cm,height=4cm]{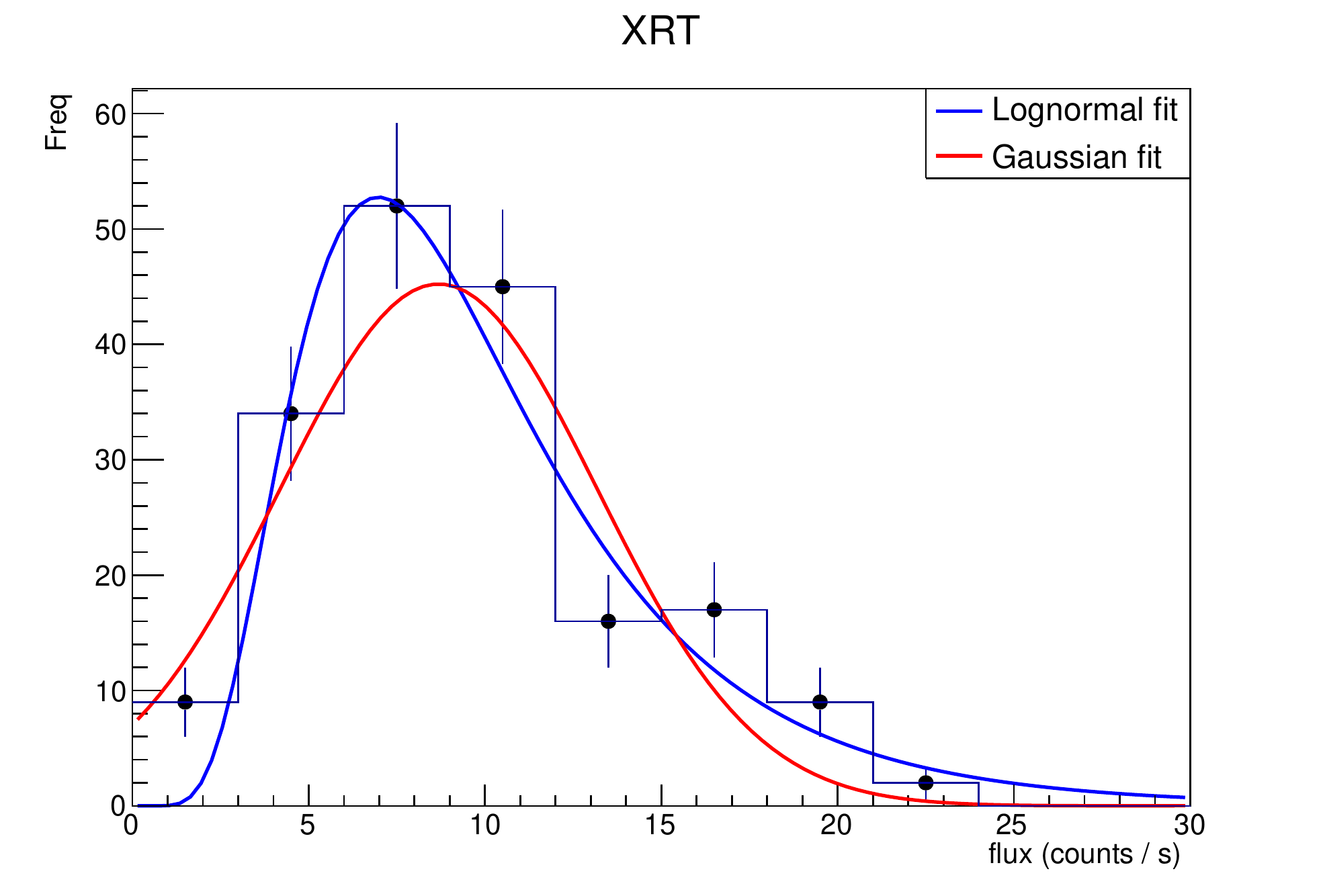} 
   \includegraphics[angle=0,width=6.5cm,height=4cm]{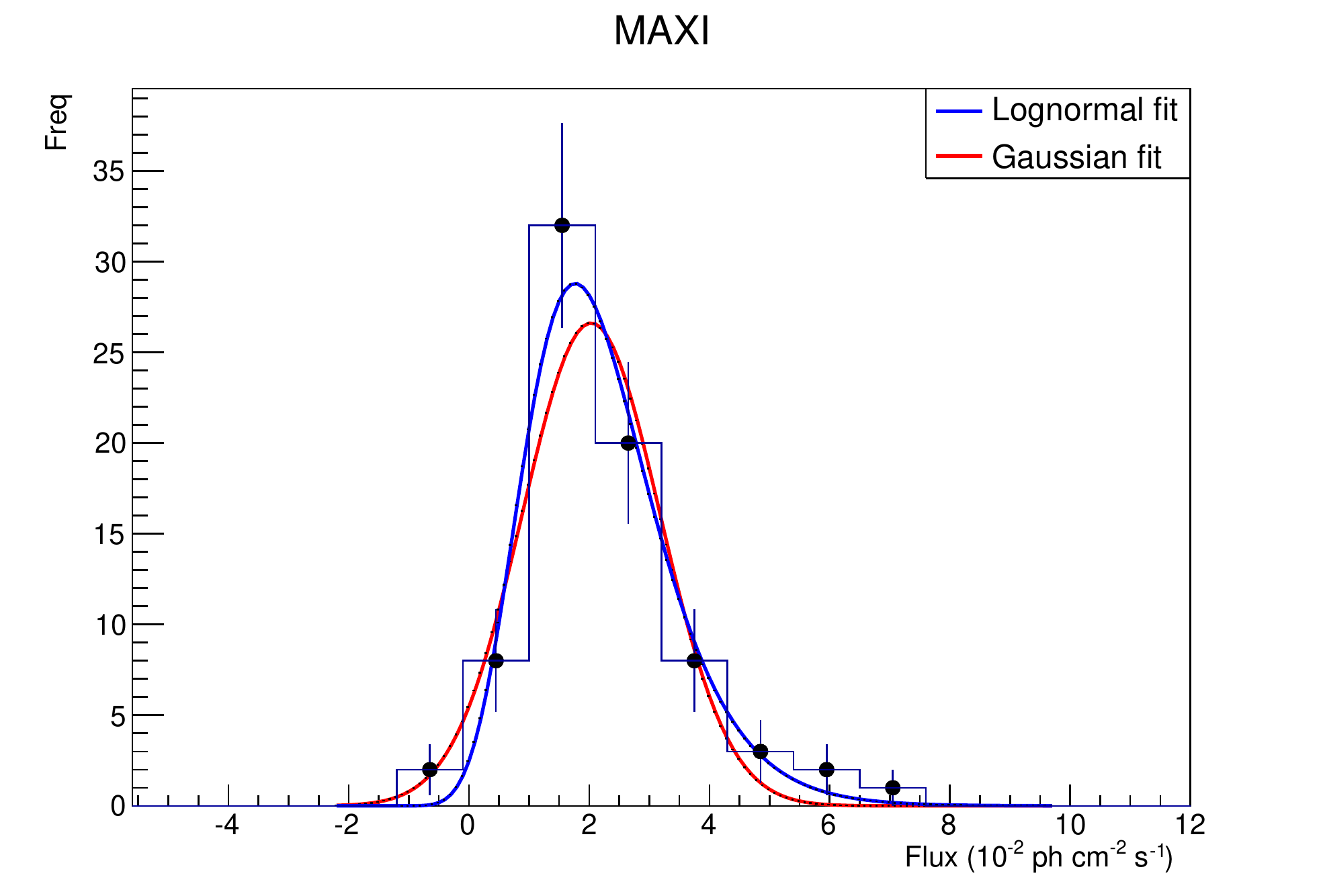} 
   \includegraphics[angle=0,width=6.5cm,height=4cm]{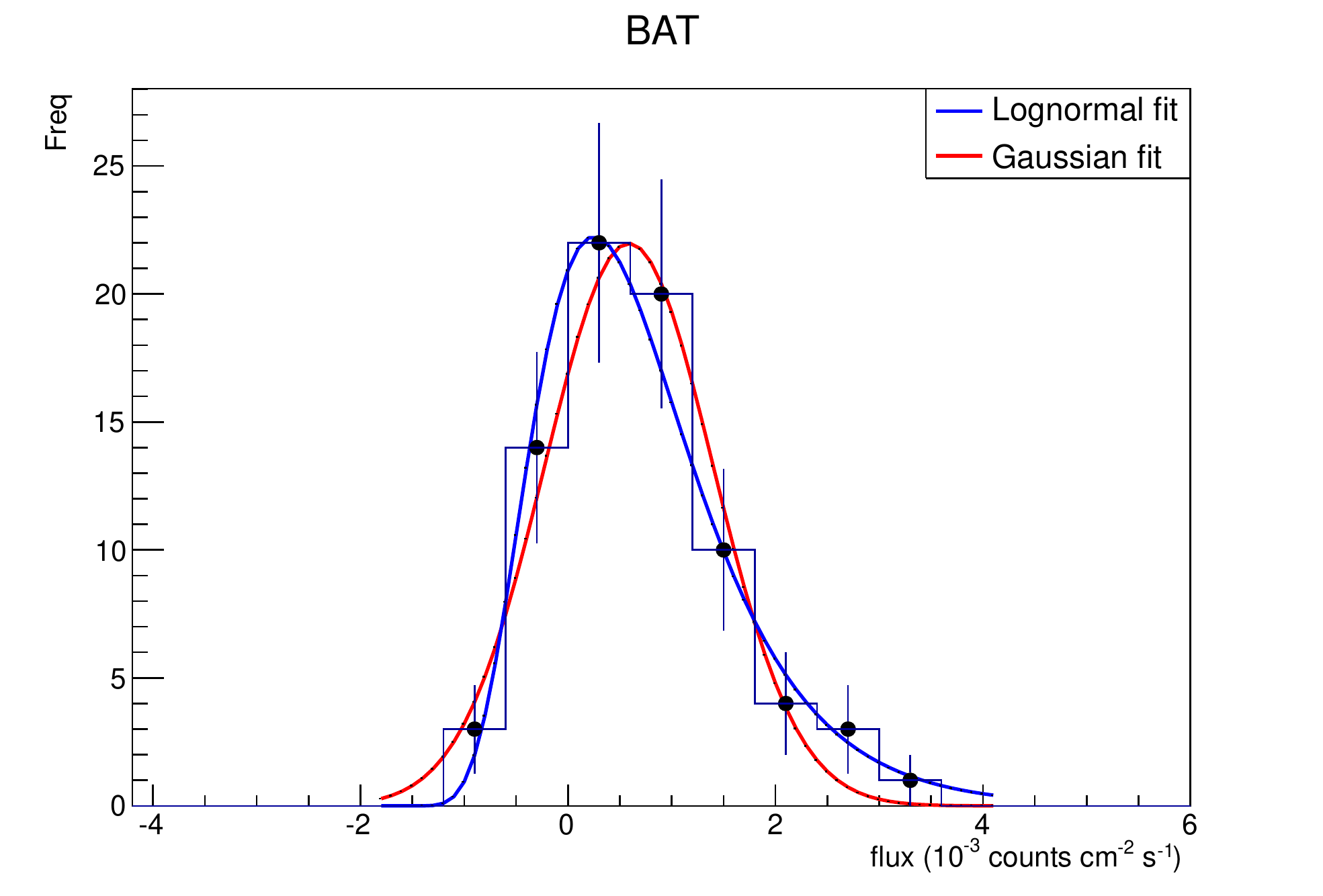}
   \includegraphics[angle=0,width=6.5cm,height=4cm]{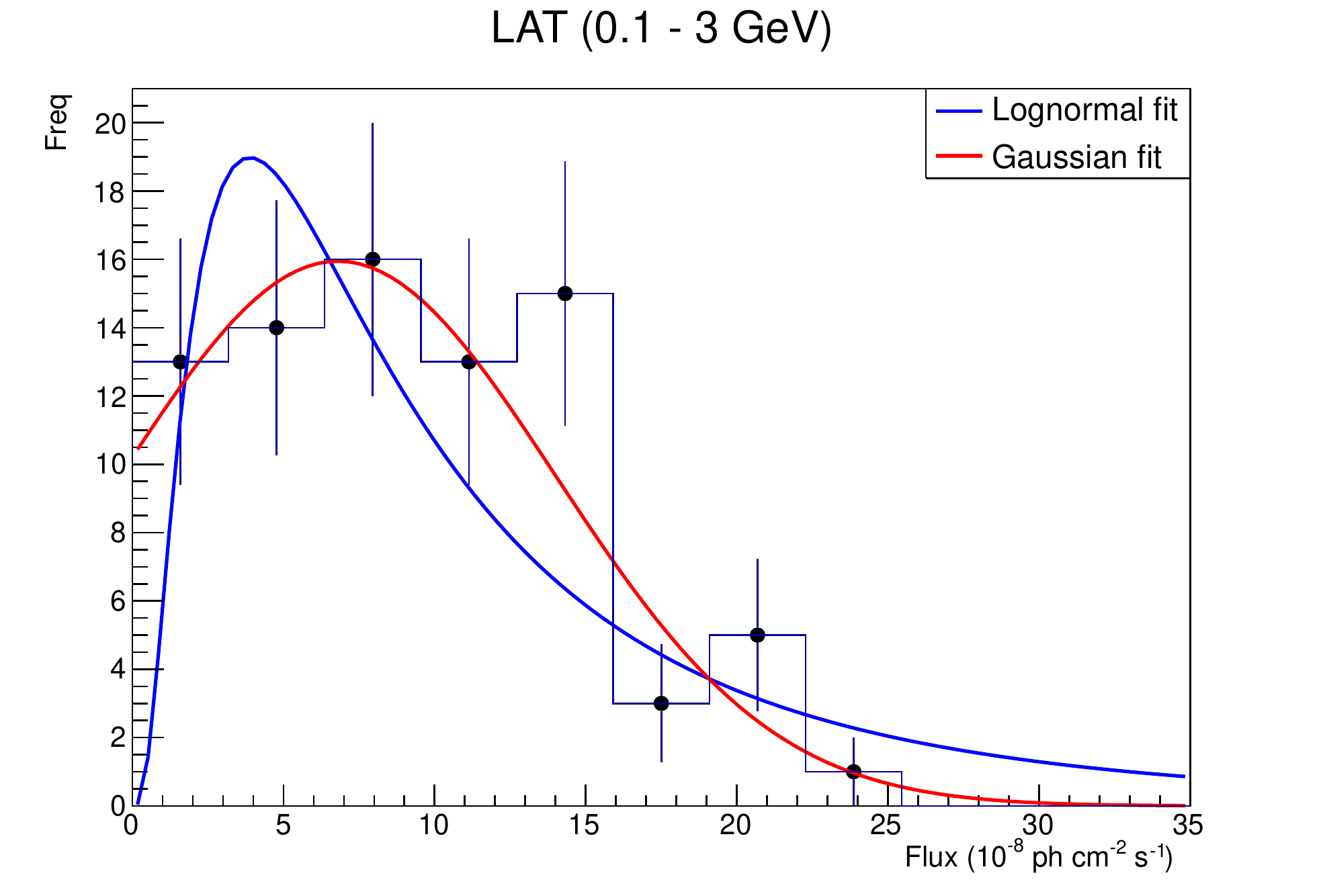} 
   \includegraphics[angle=0,width=6.5cm,height=4cm]{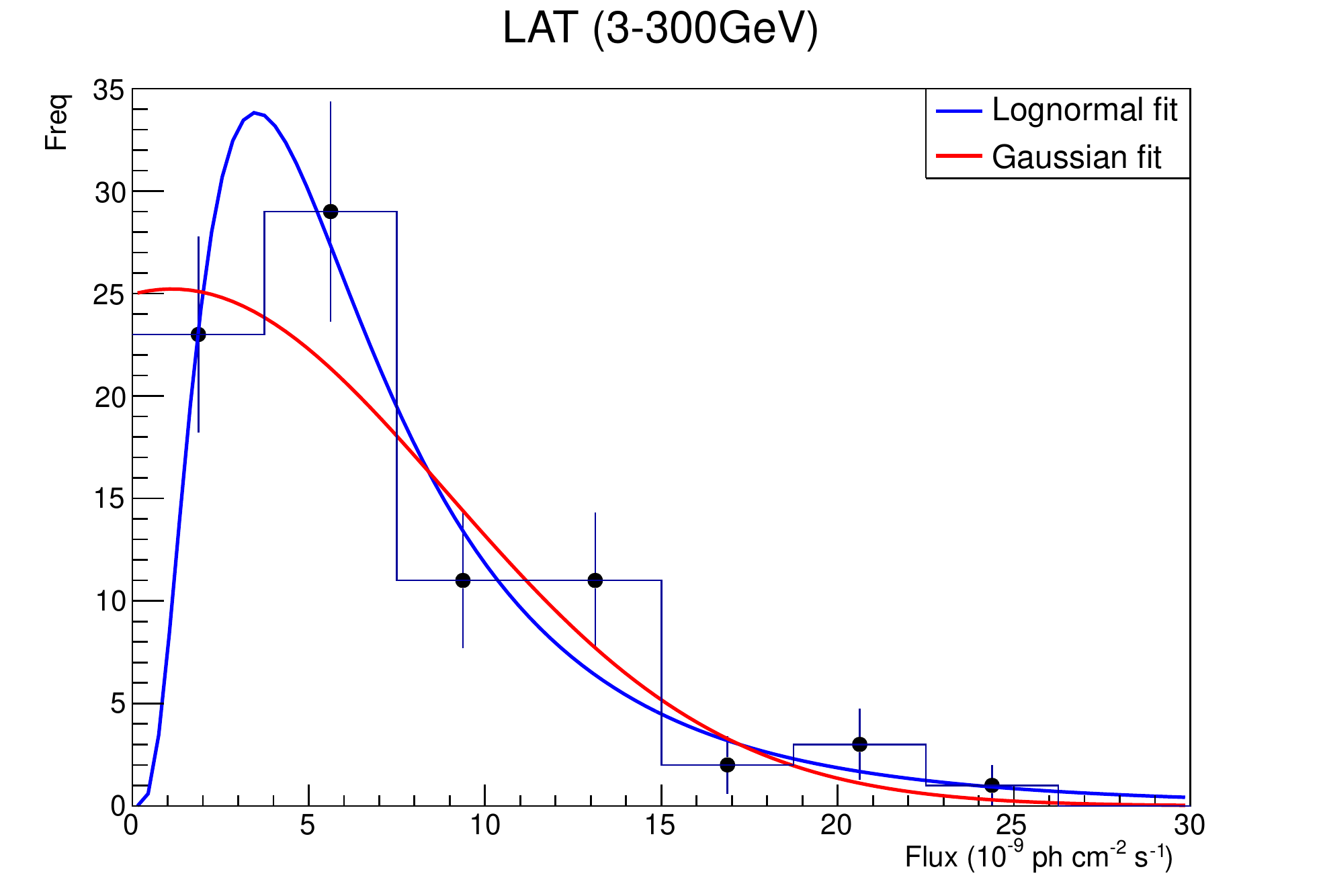}
   \caption{Gaussian and Lognormal fit to flux distribution in various energy bands. 
The binning used is same as in Fig. ~\ref{FigMWLC} except for MAXI and BAT where 10 days binning is used.}
   \label{FigDistLognorm}
\end{figure*}

\begin{figure*}[!ht]
   \centering
   \includegraphics[angle=0,width=18cm,height=4cm]{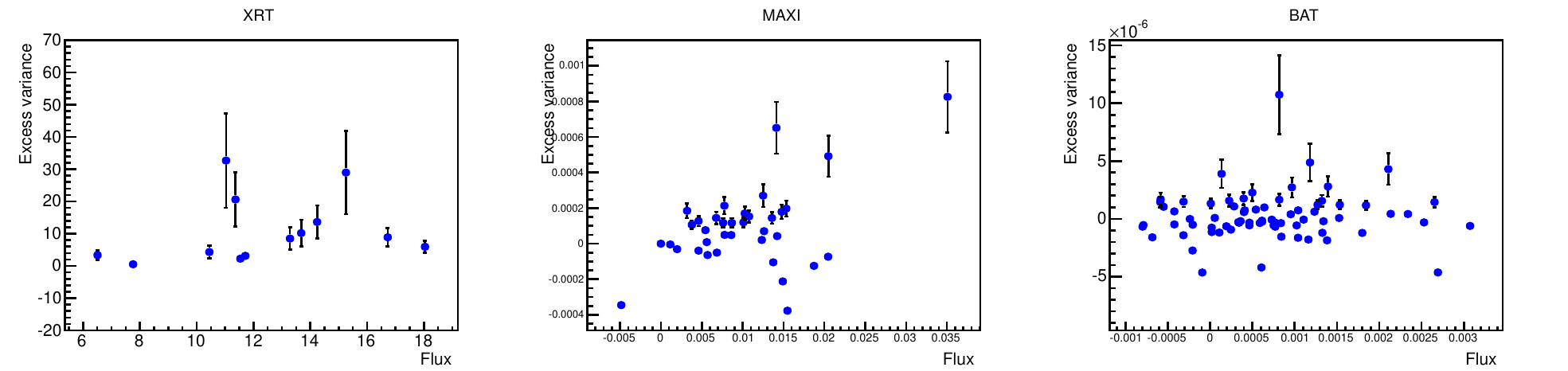}
   \caption{Excess variance vs mean flux in different X-ray energy bands : 
The data are binned into 10 days bins and excess variance is computed only 
for those bins which have at least 5 flux measurements}
   \label{rms_flux}
\end{figure*}

Lognormality, i.e., log-normal distribution of flux and linear rms-flux relation,
has been  detected in several X-ray binaries \citep{uttley, scaringi}.
It has also been detected in blazars including BL Lac (detected in X-ray regime, 
\citet{giebels}), in multiple wavebands in PKS 2155-304 \citep{chevalier},
Mkn 421 \citep{sinha16}, 1ES 1011+496 \citep{sinha17}, PKS 1510-089 \citep{kushwaha} 
etc. Flux distributions for 1ES 1959+650 covering two years' data for various
wavebands from radio to $\gamma-$ray are shown in Fig.~\ref{FigDistLognorm}. 
Data are fitted with Gaussian as well as lognormal distribution and fits are 
shown. Reduced $\chi^2$ values for both the distributions are listed in 
Table~\ref{lognorm_tab}. 
Reduced $\chi^2$ is lower for lognormal compared to Gaussian fit in all
the cases, except for 0.1-3 GeV Fermi-LAT. In order to check significance of this reduction
in $\chi^2$, the F-statistic was used. Assuming null hypothesis, i.e. no significant difference 
in variances from these two models, F-values were calculated ($F_{calculated}$).
These were compared with F-values for 95\% confidence level ($F_{95\%} $), 
for $\nu_g$ and $\nu_l$ degrees of freedom (corresponding to Gaussian and lognormal fit respectively).
These values are listed in Table~\ref{lognorm_tab} for each waveband. It can be seen that
$F_{calculated}$ are less than $F_{95\%} $ in all the cases, which means that we
can not reject null hypothesis. In other words, even though lognormal gives 
lower $\chi^2$ and better fit, Gaussian behaviour can not be ruled out at 95\% 
confidence because of low number of degrees of freedom (dof).   
Fig. ~\ref{rms_flux} shows the plot of excess variance $\sigma^2_{excess}=(S^2-\sigma^2_{err})$ as 
a function of flux for various X-ray wavebands, binned over a period of 10 days. The 
excess variance were calculated for those bins which are having at least 5 flux 
points and it is plotted against mean flux. We do not see very clear linear 
trend in these plots as was seen in case of Mkn 421 by \citet{sinha16}.

\begin{table*}
\centering
\caption{Goodness of fit for flux distributions in various energy bands}
\begin{tabular}{ccccc}   
\hline
Waveband                 & Gaussian                & Lognormal    & $F_{calculated}$  & $ F_{95\%}$  \\
                         & $\chi^2_{r}$ (dof ($\nu_g$)) & $\chi^2_{r}$ (dof ($\nu_l$)) &  &    \\
\hline
OVRO (15GHz)             & 10.82 (6) & 6.16 (6) & 1.77  & 4.28 \\
UVOT (Optical)           & 3.35 (6) & 1.80 (6)  & 1.86  & 4.28 \\
UVOT (UV)                & 5.56 (3) & 2.67 (3)  & 2.08  & 9.28 \\
Swift XRT (0.3 - 8 keV)  & 2.52 (5) & 2.18 (5)  & 1.16  & 5.05 \\
MAXI (0.2 - 20 keV)      & 6.82 (5) & 4.25 (5)  & 1.60  & 5.05 \\
Swift BAT (15 - 50 keV)  & 3.58 (5) & 1.43 (5)  & 1.60  & 5.05 \\
Fermi LAT (0.1 - 3 GeV)    & 1.08 (5) & 2.20 (5)& 2.05  & 5.05 \\
Fermi LAT (3 - 300 GeV)    & 1.69 (4) & 0.96 (4)& 1.76  & 6.39 \\
\hline
\end{tabular}

\label{lognorm_tab}
\tiny{
\tablefoot{dof is number of bins minus number of model parameters which is three for both the models; 
$ F_{calculated}$ is ratio of two variances with larger variance in numerator ;
 $F_{95\%} $ is F-statistic value at 95\% confidence level for $\nu_g$ and $\nu_l$ degrees of freedom
}}

\end{table*}

\subsection{Spectral studies}

We studied X-ray spectra using Swift-XRT data for the flare state during MJD 57530 - 57589. 
This flare was divided into six states, each spanning ten days period. These states 
are marked by vertical dotted lines in Fig.~\ref{FigMWLC}.
We also studied X-ray spectrum in the interval MJD 57177 - 57187 which corresponds
to the quiescent state of the source. These seven average spectra for each bin were fitted with a powerlaw 
with line of sight absorption, over the energy range of 0.3-8 keV. 
Alternatively spectra were also fitted with logparabola model with line of sight absorption. 
Best fit parameters for both the models are given in Table~\ref{tabXrayFit}. 
Logparabola seems to fit data better than powerlaw, as seen 
from improvement in values of reduced $\chi^2$.

\begin{figure}[!ht]
   \centering
   \includegraphics[angle=0,width=9cm]{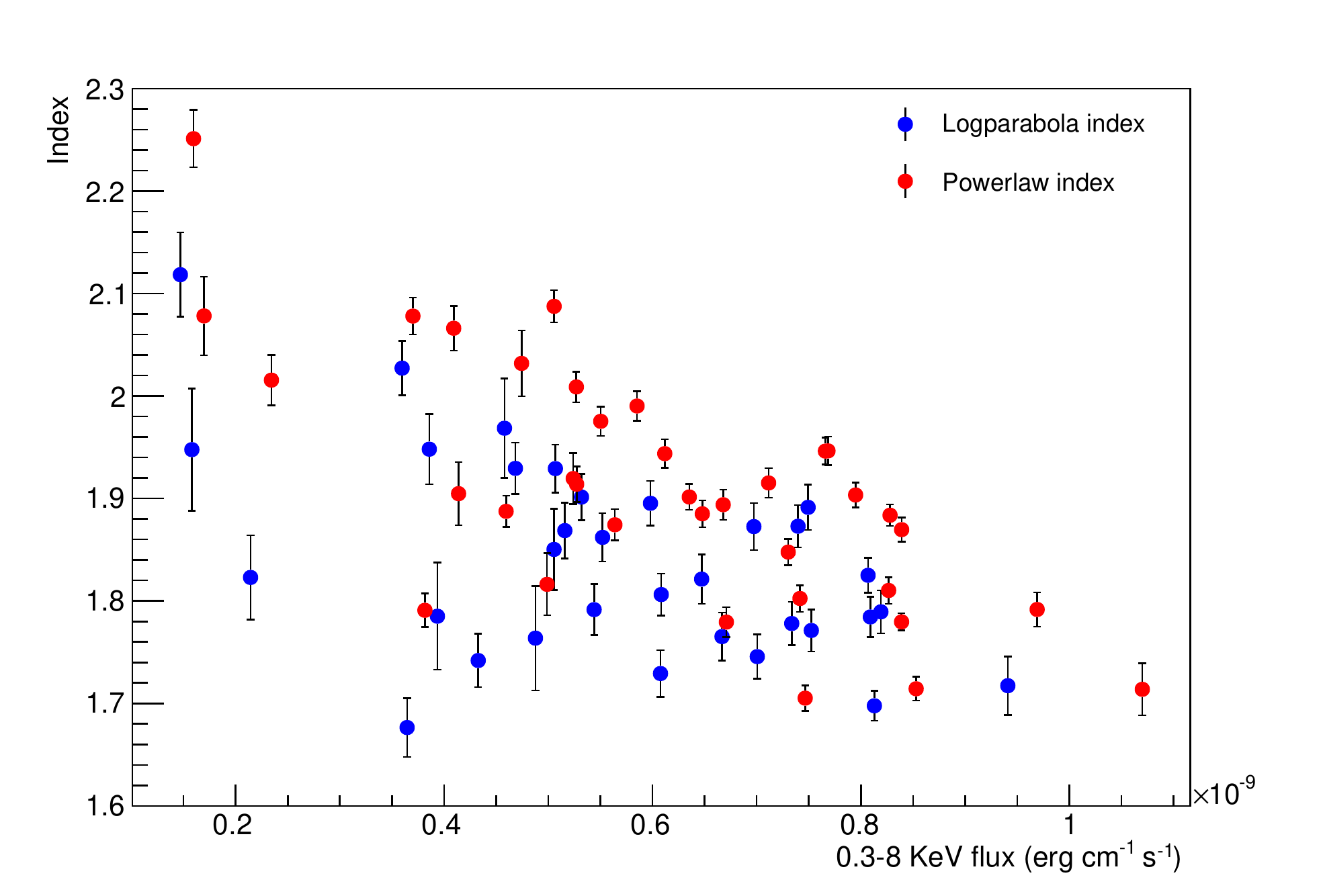}            
   \caption{Red: Powerlaw index vs X-ray flux (The correlation coefficient is -0.72), Blue: Logparabola index vs X-ray flux (The correlation coefficient is -0.61) }
   \label{FigXRTIndFlux}
\end{figure}

\begin{figure}[!ht]
   \centering
   \includegraphics[angle=0,width=9cm]{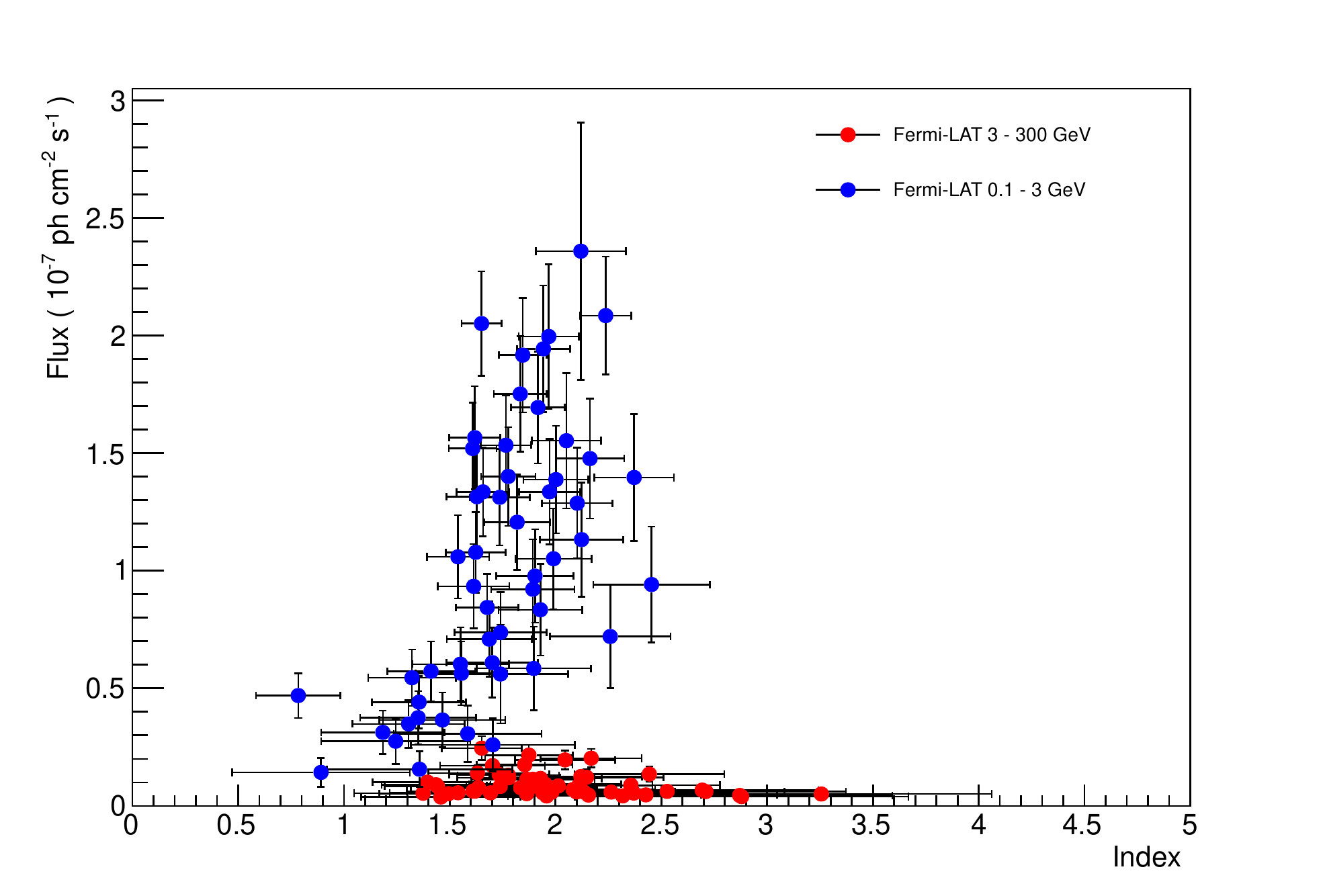}
   \caption{Powerlaw index vs LAT flux. The correlation coefficient for
0.1 - 3 GeV band is 0.60 }
   \label{FigLATIndFlux}
\end{figure}

\begin{figure}[!ht]
   \centering
   \includegraphics[angle=0,width=9cm]{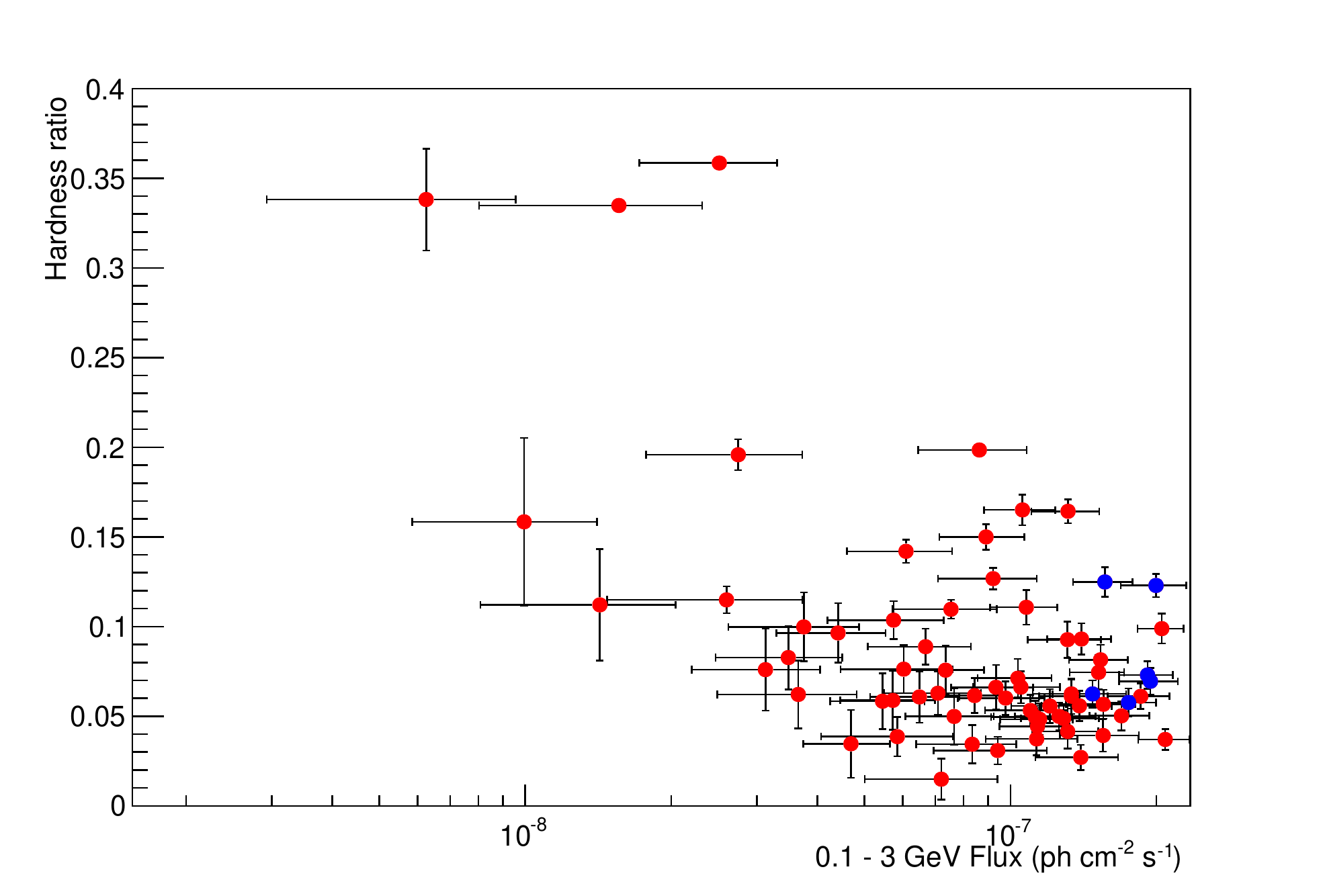}
   \caption{Hardness ratio in Fermi band vs 0.1 - 3 GeV flux, Points corresponding to flare is shown in blue}
   \label{FigLATHard}
\end{figure}

\begin{table*}
\centering
\caption{Spectral fitting parameters for 10 days binned Swift-XRT data}
\begin{tabular}{cccccc}
\hline

Period       &  Poweraw           & Reduced $\chi^2$ & \multicolumn{2}{c}{Logparabola}   & Reduced $\chi^2$ \\
(MJD)        &  Index              &(prob)   & $\alpha$        & $\beta$           &       (prob)           \\         
\hline
Q (57177-57186)    &       2.19 +/- 0.02 &  1.34 (1.55E-03)      &              2.04 +/- 0.00 &  0.49 +/- 0.07  & 1.05 (2.91E-01)              \\
F1 (57530-57539)   &       2.09 +/- 0.01 &  1.44 (3.03E-07)     &              1.97 +/- 0.02 &  0.33 +/- 0.04  & 1.17  (2.16E-02)           \\
F2 (57540-57549)   &       1.89 +/- 0.01 &  1.72 (2.92E-27)     &              1.81 +/- 0.01 &  0.21 +/- 0.01  & 1.21 (1.89E-04)             \\
F3 (57550-57559)   &       1.82 +/- 0.01 &  1.58 (2.16E-19)    &              1.75 +/- 0.01 &  0.15 +/- 0.01  & 1.33 (2.56E-08)             \\
F4 (57560-57569)   &       1.88 +/- 0.02 &  0.96 (6.81E-01)     &              1.81 +/- 0.02 &  0.17 +/- 0.04  & 0.91  (8.65E-01)            \\
F5 (57570-57579)   &       1.74 +/- 0.01 &  1.28 (6.81E-05)     &              1.61 +/- 0.02 &  0.28 +/- 0.03  & 1.05  (2.05E-01)            \\
F6 (57580-57589)   &       1.96 +/- 0.01 &  1.39 (1.51E-08)     &              1.87 +/- 0.01 &  0.23 +/- 0.02  & 1.13  (2.64E-02)            \\
\hline
\end{tabular}

\vspace{0.1cm}

\tablefoot{Values in the brackets in column 3 and 6 denote the value of null hypothesis probability.}

\label{tabXrayFit}
\end{table*}

We also analyzed X-ray spectra from individual observations of XRT during the flare.
Fig.~\ref{FigXRTIndFlux} shows the plot of flux in 0.3-8 keV range as a function
of X-ray photon index for both powerlaw and logparabola model.
X-ray spectrum seems to harden with the increase in the
flux. Similar behaviour from this source is also reported by \citet{indexflux}.
However no such correlation was found when source was observed in low state
during 2007 - 2011 \citep{aliu2013}. Spectral hardening with increase in
flux is seen in several other blazars including Mkn 421 \citep{sinha16}; Mkn 501 \citep{kraw}.

We also investigated spectral parameters in $\gamma-$ray band using Fermi-LAT data,
dividing it into two energy ranges, 0.1-3 GeV and 3-300 GeV. Spectra binned over 
ten days were fitted with a powerlaw. Variation of flux with powerlaw index for
both low and high energy bands is shown in Fig.~\ref{FigLATIndFlux}. While lower
energy band shows increase in the flux with increase in the photon index, no such
trend is seen at higher energies. Similar behaviour at $\gamma-$ray
energies was seen for Mkn 501 \citep{amit}. Such trend might be due to different
emission regions for low and high energy $\gamma$-rays.
Fig.~\ref{FigLATHard} shows hardness ratio, defined
as a ratio of 3-300 GeV to 0.1-3 GeV flux, as a  function of 0.1-3 GeV flux. 
This also indicates that for the period of 800 days the spectrum is 
becoming softer at higher flux.

\section{SED evolution and modeling}

We have studied the evolution of SED during flare. For this purpose, multiwaveband
SEDs were generated for six flux states in the flare as well as for one quiescent
state. First we tried to model these SEDs using simple single zone Synchrotron 
Self Compton (SSC) model using the code developed by \citet{orpkra}. This model 
assumes the emission zone to be a blob of radius R traveling down the jet with 
bulk Lorentz  factor $\Gamma $ towards the observer at an angle of $\theta $. 
The emission zone is filled with non-thermal electron distribution and randomly 
oriented magnetic field.  The electron population can be described by broken 
powerlaw, having low and high energy indices $p_1$ and $p_2$.
The radius R of the emission region can be constrained by observed doubling time scale $t_{var}$ 
using relation,
\begin{equation}
 R \sim \frac{c \delta t_{var}}{(1+z)} 
\end{equation}

The six flux states during different epochs of the flare are denoted by F1, F2, F3,F4, F5 and 
F6 while quiescent state is denoted by Q. The SEDs were modeled with jet 
parameters $\theta$ and $\Gamma$ to be 3$^\circ$ and 9.4 respectively which corresponds to
Doppler factor of $\sim$ 15. The value of Doppler factor is consistent with
the value reported in the past \citep{orpkra,indexflux}.
The blob radius of 7.84 $\times$ $10^{16}$ cm was assumed.
While modeling $\theta$ is kept fixed at 3 degree and it is also assumed 
that emitting region has same cross section as jet.

We find that due to sharp curvature of SEDs in X-ray region, we always 
underpredict the optical/UV flux with one zone SSC model. Also except for F2 
state, one zone SSC model does not fit $\gamma$-ray flux below 
$\sim$ 3 GeV. Hence we made an attemp to explain the observed SEDs with two 
zone SSC model in which resultant emission is the sum of emission from two 
comoving blobs of different sizes. We assumed the size of the second (outer)  
blob to correspond to $t_{var}$  $\sim$ 10 days and having same Doppler factor 
as that of the inner blob.
The cross plot of the flux-power law index in 0.1-3 GeV and 3-300 GeV bands 
suggests the different emission regions for these energy bands 
(Fig.~\ref{FigLATIndFlux}).  It is reported in long term (2005-2014) 
multiwavelength cross-correlation study that one zone SSC scenario was not 
always suitable to explain the emission from this source \citep{opt}.

\begin{figure*}
\centering
  \includegraphics[angle=0,width=7cm]{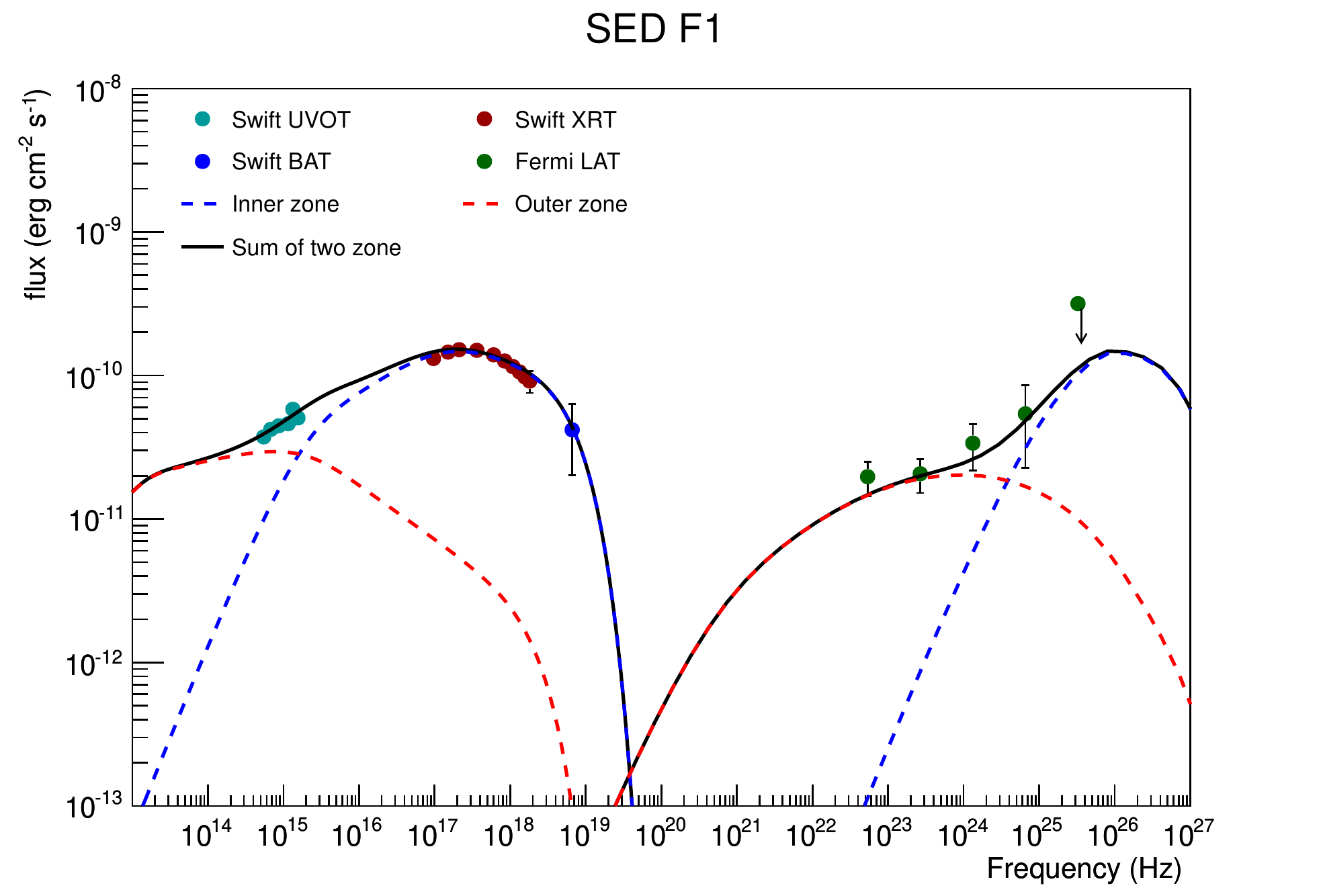}
  \includegraphics[angle=0,width=7cm]{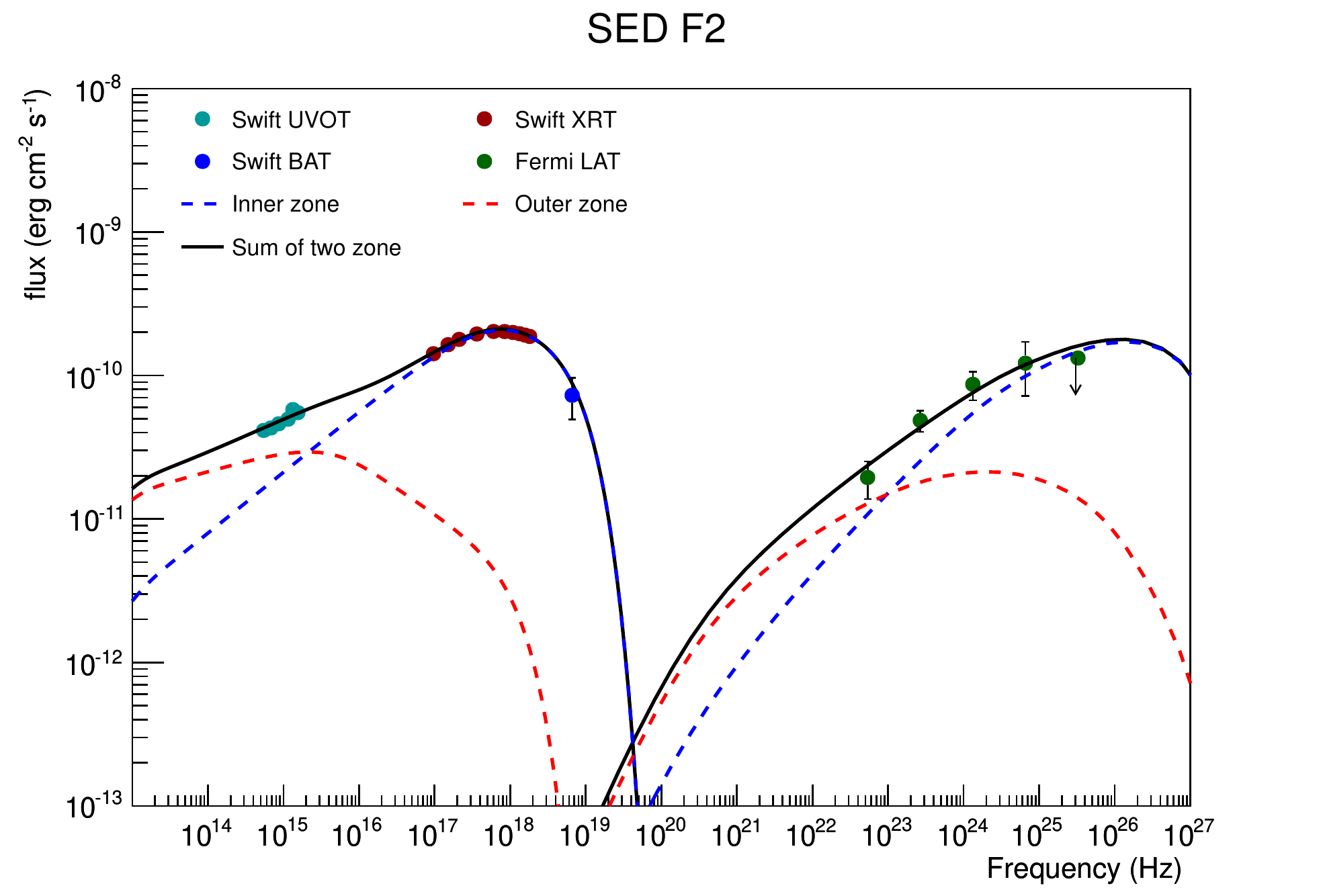}
  \includegraphics[angle=0,width=7cm]{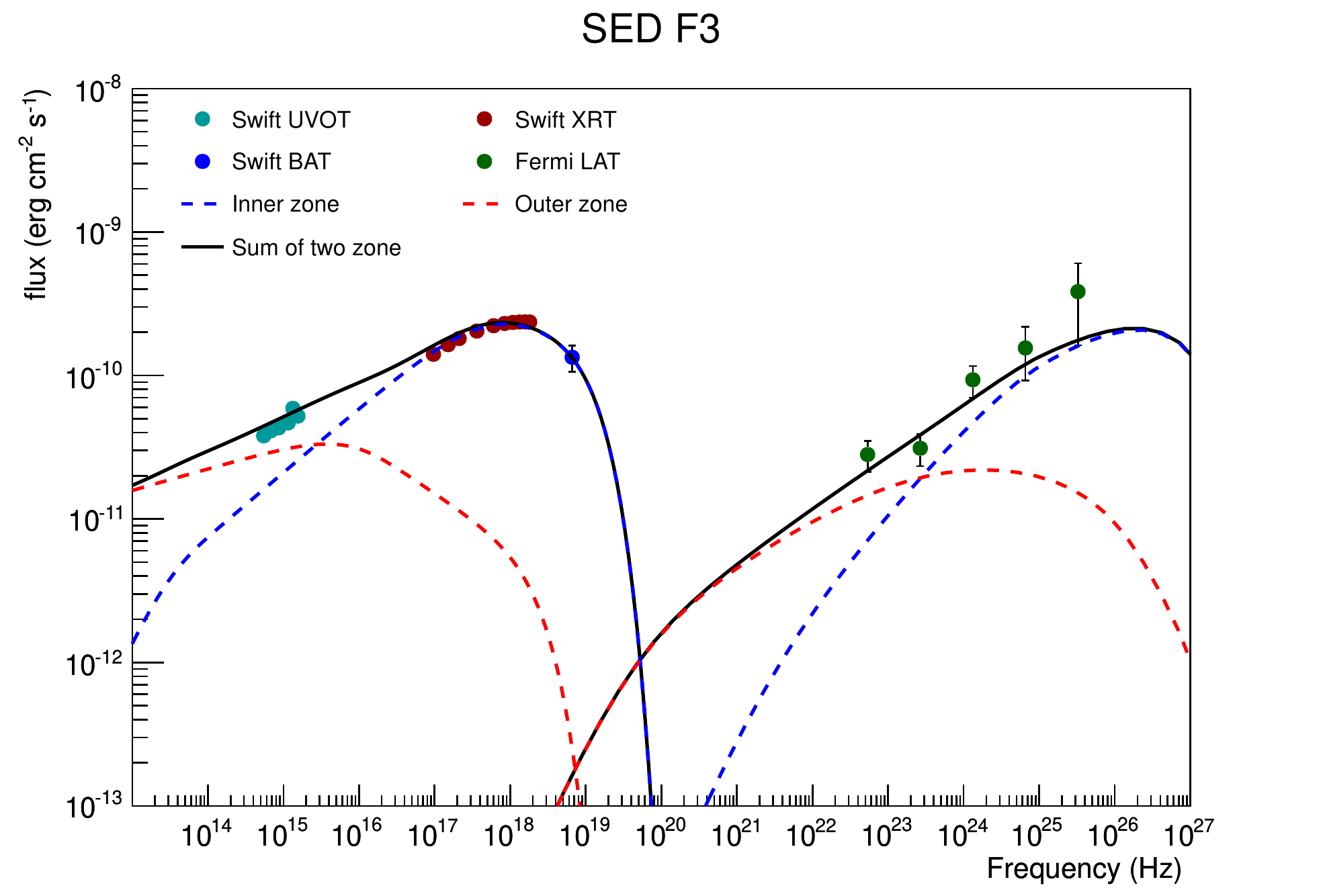}
  \includegraphics[angle=0,width=7cm]{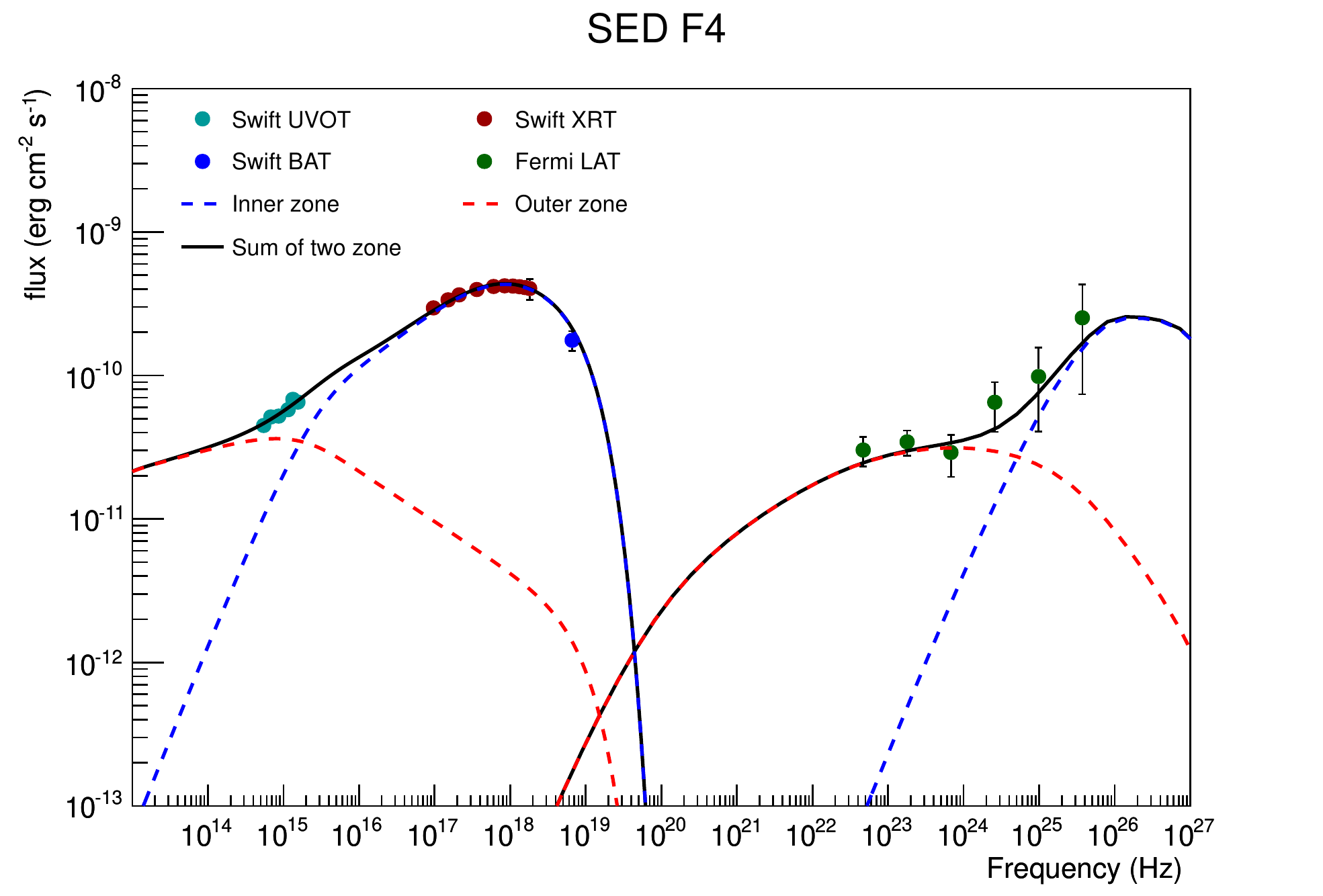}
  \includegraphics[angle=0,width=7cm]{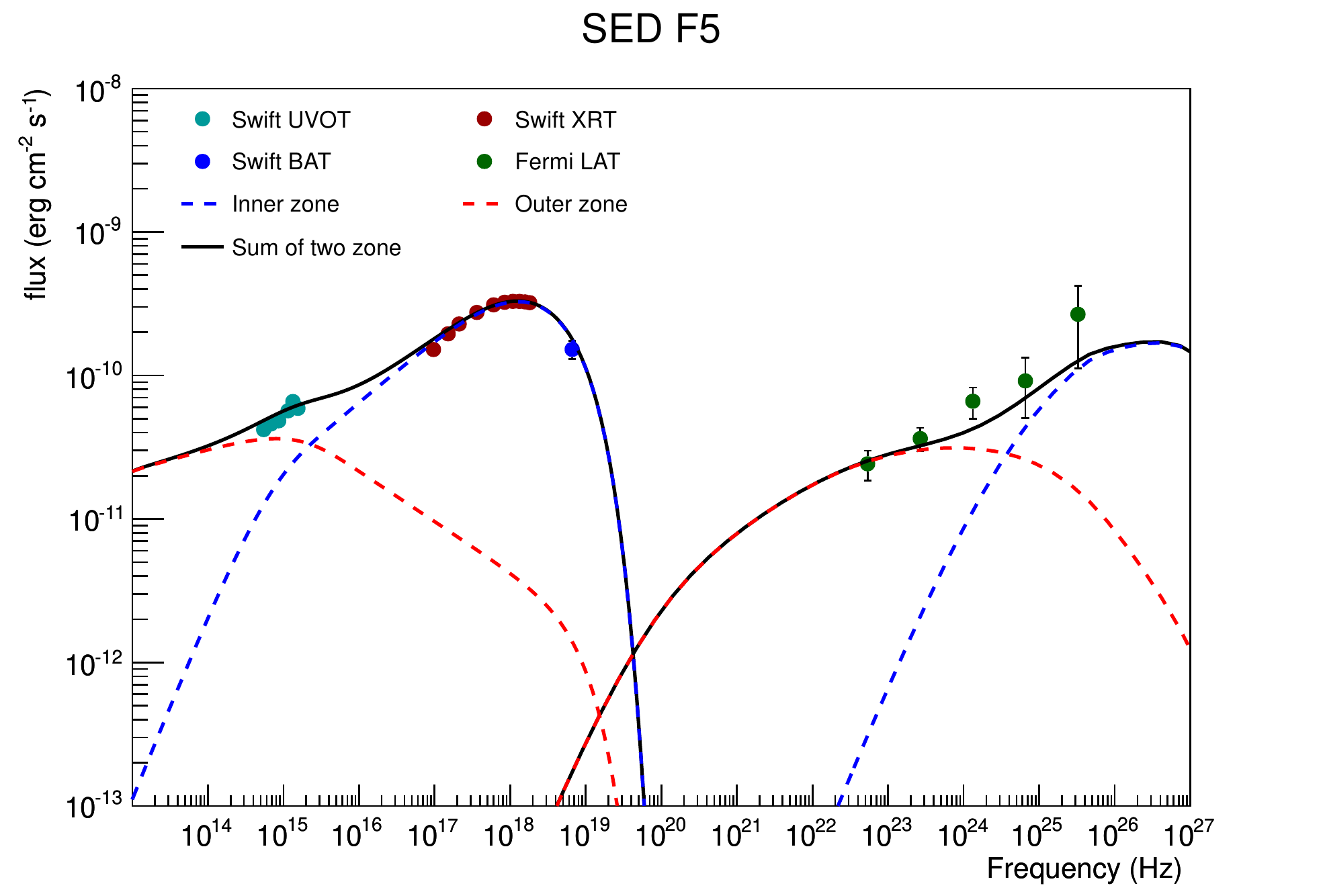}
  \includegraphics[angle=0,width=7cm]{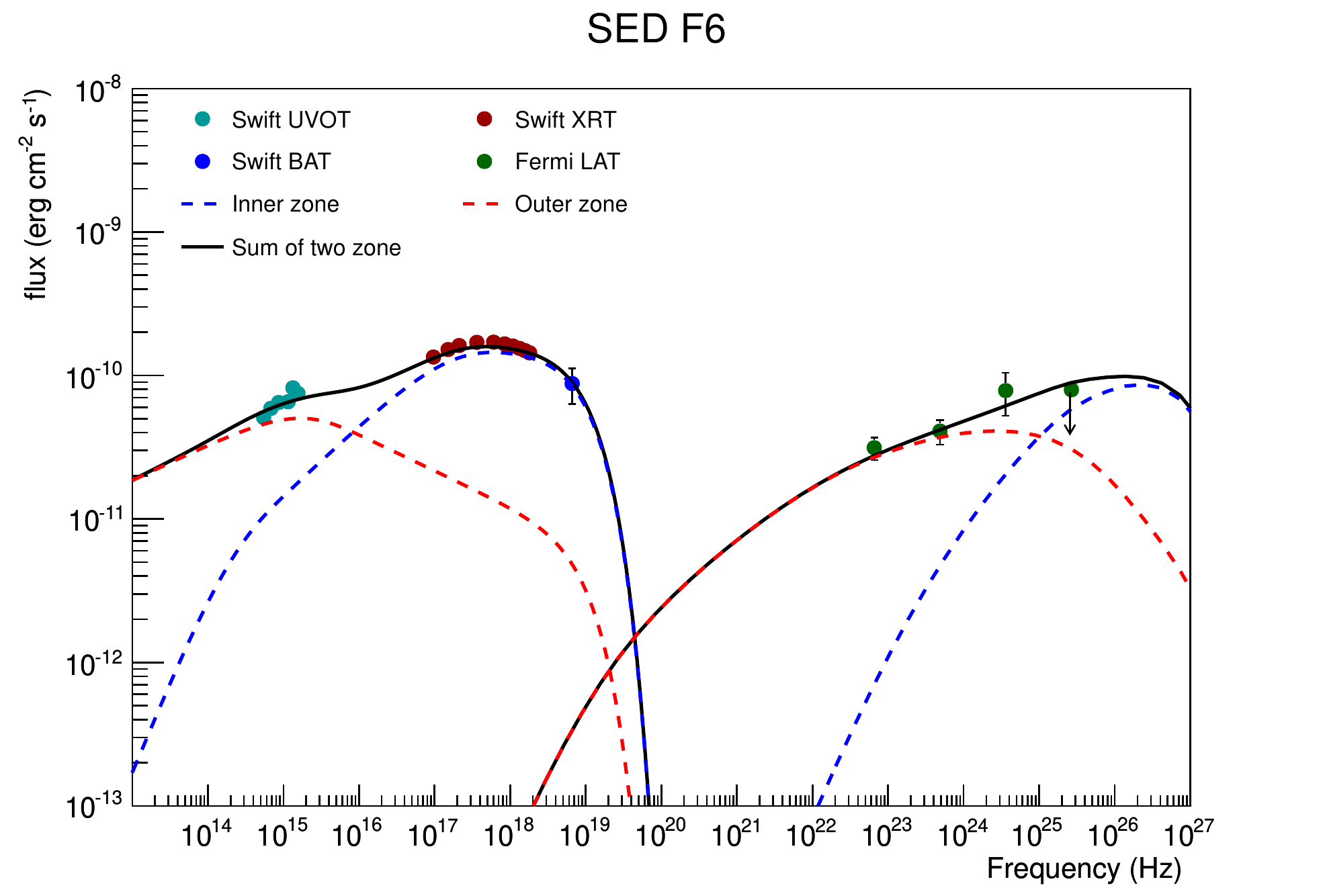}
  \caption{SEDs fitted with two zone SSC model for six states during flare}
  \label{FigSED}
\end{figure*}

\begin{figure}[!ht]
\centering
  \includegraphics[angle=0,width=9cm]{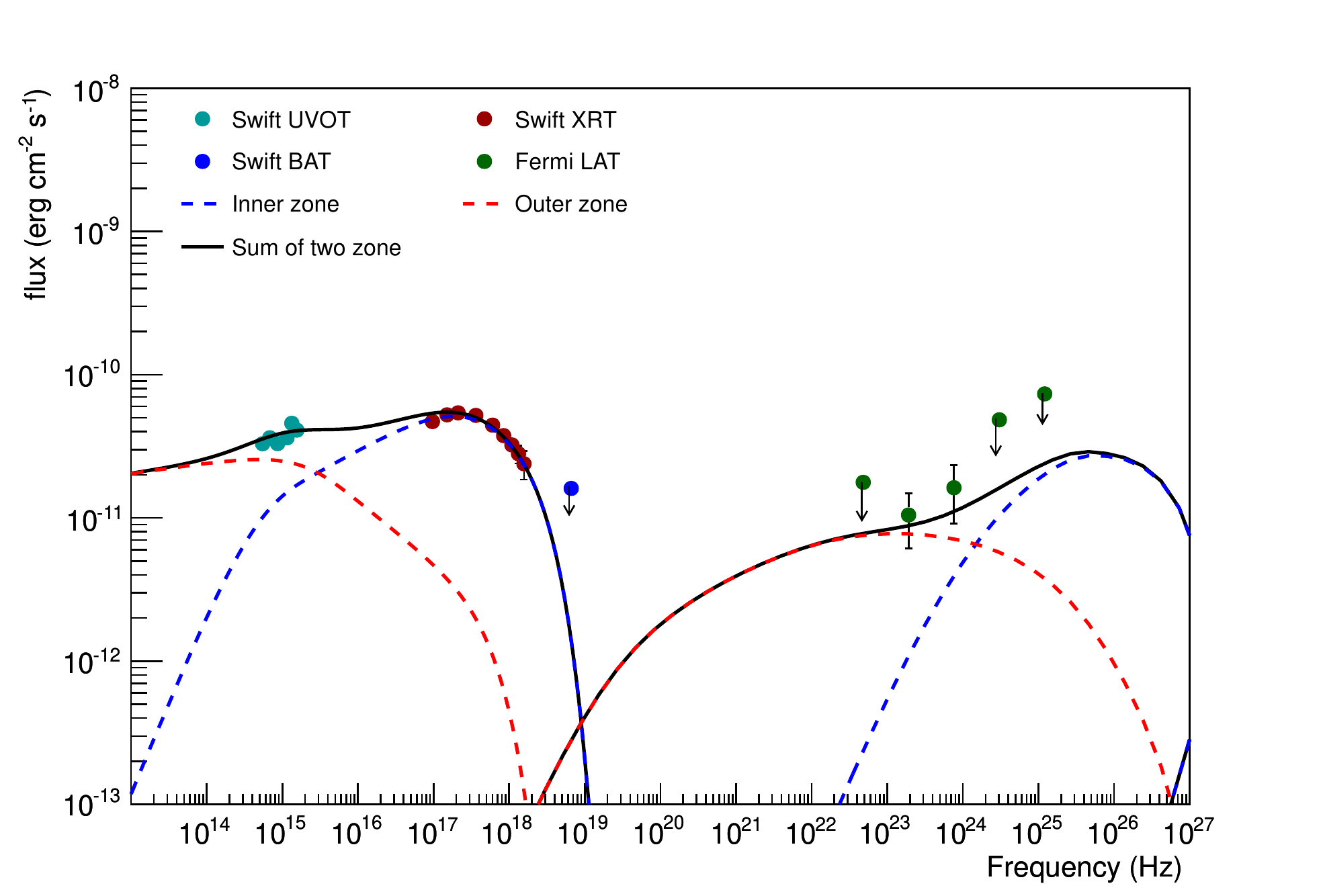}
  \caption{Quiescent state SED fitted with two zone SSC model}
  \label{FigQ}
\end{figure}

\begin{figure}[!ht]
\centering
  \includegraphics[angle=0,width=9cm]{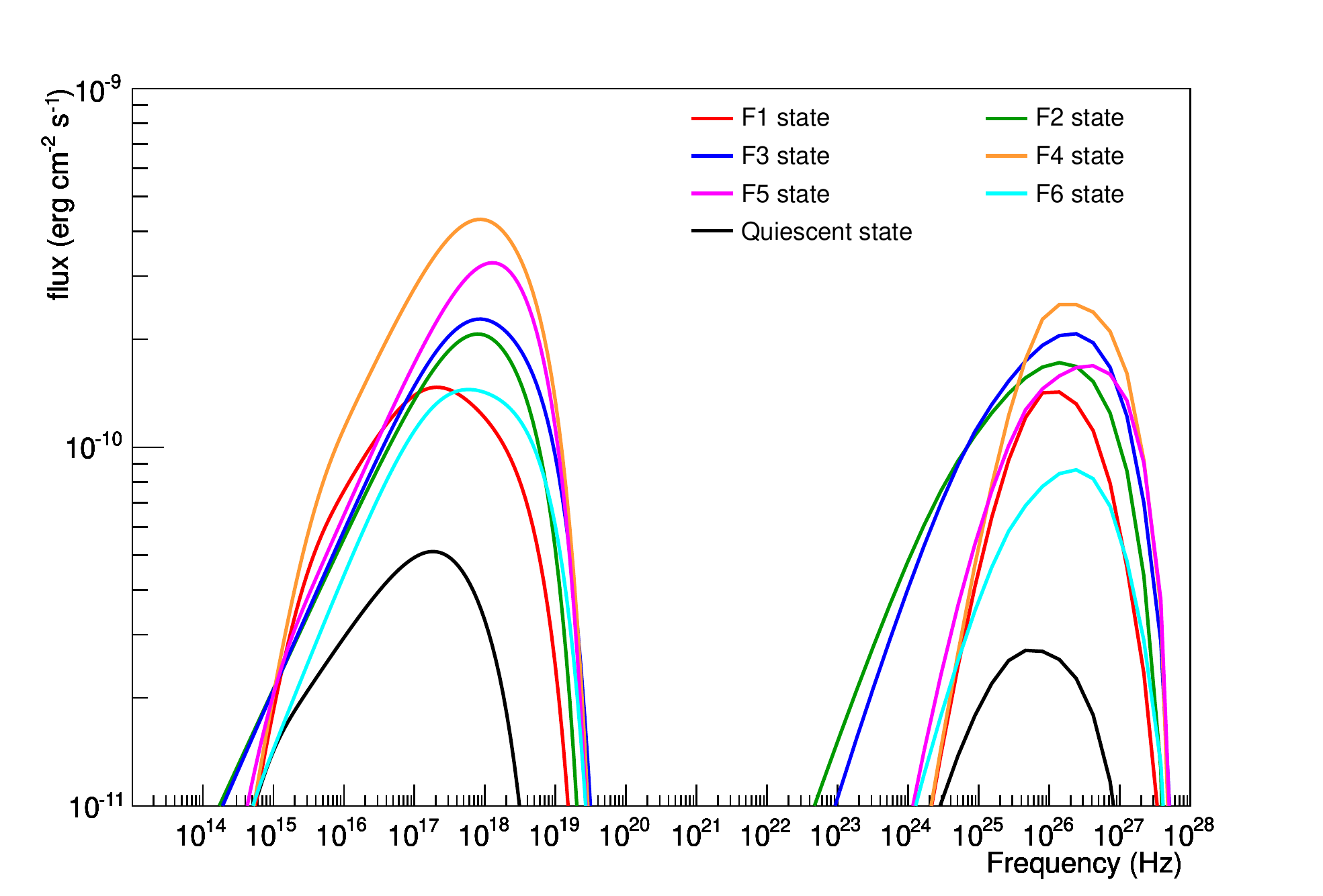}
  \caption{Variation of inner zone of SSC emission during 6 flare and one quiescent states}
  \label{Figall6}
\end{figure}

The fit for all six high states is shown in Fig.~\ref{FigSED} and one 
quiescent state is shown in Fig.~\ref{FigQ}. The
Fig.~\ref{Figall6} shows the variation of inner blob emission as the flare 
evolves. The SSC model parameters for two zones are listed in Table.~\ref{SED}.
 Apart from conventional single-zone SSC model, External Compton 
(EC), lepto-hadronic model and two independent SSC models were tried 
for fitting SEDs in the past \citep{2zone}. Authors report that two 
zone SSC model described the observed SED reasonably well for this source. 
The two zone SSC model was also used to reproduce observed SED during 
2011 observation of Mkn 501 \citep{amit, amit2016}.

\begin{table*}
\centering
\caption{SSC model parameters for flare and quiescent states for two zones with 
Doppler factor ($\delta$) $\sim$ 15, $\theta$ = 3, $R_{inner}$=7.84 $\times 10^{16}$ cm ($t_{var} \sim$ 2 days) 
and $R_{outer}$=3.92 $\times 10^{17}$ cm ($t_{var} \sim$ 10 days)}

\resizebox{\textwidth}{!}{
\begin{tabular}{ccccccccc}
        \hline
        SED state & B [10$^{-2}$ G] & $U_{e}$ [10$^{-3}$ erg/cc]  & $\gamma_{min} $ [10$^{4}$] & $ \gamma_{max}$ [10$^{6}$] & $ \gamma_{br}$ [10$^{5}$]  &$ p_1$ & $p_2$ & $\eta$ ($U_e$/$U_B$)\\
        \hline
        Q (Inner blob) & 1.10 & 0.50 & 2.46 & 1.55 & 5.52 &2.40 & 3.40 & 103.85     \\
        F1 (Inner blob)& 0.95 & 1.10 & 5.51 & 3.10 & 4.92 &2.35 & 3.35 & 306.33  \\
        F2 (Inner blob)& 1.01 & 1.85 & 0.31 & 3.10 & 9.81 & 2.15& 3.15 & 451.32   \\
        F3 (Inner blob)& 0.93 & 1.80 & 0.62 & 3.90 & 9.37 & 2.10 & 3.10& 528.72  \\
        F4 (Inner blob)& 1.05 & 1.50 & 6.18 & 3.32 & 9.81 & 2.15 & 3.15& 341.92   \\
        F5 (Inner blob)& 0.90 & 1.40 & 3.47 & 3.48 & 14.18& 2.10 & 3.10& 434.39   \\
        F6 (Inner blob)& 0.90 & 1.00 & 1.96 & 3.90 & 6.19 & 2.00 & 3.00& 341.31   \\
        \hline
        Q (Outer blob) & 0.70  & 0.13 & 0.20 & 0.98 & 0.62 & 2.85& 3.85        & 66.07   \\
        F1 (Outer blob)& 0.45  & 0.14 & 0.49 & 2.46 & 0.78 & 2.75& 3.75        & 170.65   \\
        F2 (Outer blob)& 0.43  & 0.14 & 0.44 & 1.96 & 1.24 & 2.70& 3.70        & 194.94  \\
        F3 (Outer blob)& 0.48  & 0.18 & 0.25 & 2.46 & 1.55 & 2.70& 3.70        & 194.94    \\
        F4 (Outer blob)& 0.45  & 0.23 & 0.31 & 4.92 & 0.74 & 2.70& 3.70        &279.25    \\
        F5 (Outer blob)& 0.45  & 0.23 & 0.31 & 4.92 & 0.74 & 2.70& 3.70        &279.25   \\
        F6 (Outer blob)& 0.50  & 0.22 & 0.20 & 4.92 & 0.78 & 2.50& 3.50        &221.17    \\
        \hline
\end{tabular}
}
\vspace{0.01cm}
\tiny{
\tablefoot{ $U_{e}$ : Electron energy density ;  
 $\gamma_{min}$ : Minimum value of Lorentz factor of electron present in the emission region ;
 $\gamma_{max}$ : Maximum value of Lorentz factor of electron present in the emission region ; 
 $\gamma_{br}$ : Lorentz factor at break in electron injection spectrum ;
 $\eta$ : Equipartition coefficient} }
\label{SED}
\end{table*}

In homogeneous leptonic model, one expects a break in electron energy 
distribution (EED) where Lorentz factor is given by
\begin{equation}
\gamma_{c} =  \frac{3\pi m_{e} c^{2}}{\sigma_{t}B^{2}R}
\end{equation}
At this $\gamma_c$, the escape time from the source equals the synchrotron 
cooling time \citep{abdo,graff}. The Fig.~\ref{FigGammaB} shows 
variation of fitted B with $\gamma_{br}$ for flare states F1-F6. Theoretical 
curve based on Eq.(9) assuming R = 7.84 $\times$ 10$^{16}$ cm is shown by 
solid line for inner blob. 
As per Eq. (9), we find that observed B is upto $\sim$ 45 \% lower than the expected one (Fig.~\ref{FigGammaB}). 
This means that we observe break in electron spectrum at lower energies than
the one expected in case of similar escape and cooling time scales.

\section{Discussion and conclusions}

The lognormal fit to the flux distribution of two years of data gives improved reduced 
$\chi^2$ as compared to Gaussian fit. However we do not see this difference 
between models to be significant based on F-test. This could be due to
smaller size of the data set. Since tail is seen in the flux distribution,  
possibly longer data set may show significant preference to lognormal fit.
Several authors have seen lognormal behaviour of
flux distribution in X-ray binaries and BL-Lac sources. The lognormal flux behaviour
in BL-Lac is generally considered as indication of the accretion disk variability’s
imprint onto the jet \citep{McHardy}. 
In the present data set, we do not see lognormality convincingly.
Longer data set may lead to conclusive result.

The temporal analysis of two years data on 1ES 1959+650 shows that source did not exhibit significant
variation in optical band. However, source showed significant flux variation in X-ray and 
$\gamma-$ray band.  The DCFs are computed to quantify the correlations and lags
between measurements from various instruments covering different energy bands over 
two years. Significant correlation is seen between fluxes of X-ray and both 
high and low energy $\gamma$-rays. No significant correlation of optical/UV fluxes 
with other energy bands was seen. This could be due to different emission region of 
optical/UV photons and these photons are up-scattered to low energy $\gamma$-rays, which 
might be observed by Fermi-LAT.

We found the X-ray spectrum to be curved during most of the observations and hence logparabola 
fit to the X-ray spectrum is found to be better than conventional power law fit. 
Log parabola fit to X-ray data also suggest that particle acceleration should be stochastic in nature.
The spectral evolution was found to be 'harder when brighter'in X-ray during the observation 
period of $\sim$ 2 months covering flare. However, 'softer when brighter' behaviour was also seen 
in $\gamma$-rays (Fig.~\ref{FigLATHard}) when source is not very active.

We found a significant change in  electron density across the flare in both inner and
outer regions by modeling of SEDs. However a maximum $\sim$ 20 $\%$ change is observed in magnetic field. 
Either injection of new particles or re-acceleration of particles might have increased 
electron energy density in the inner zone. This enhanced electron energy density might be 
responsible for the high state and flaring activity of the source.  
Also similar magnetic field in outer blob might be due to re-amplification of
magnetic filed by passing shock wave.
Narrow EED and  spectral hardening are found in the electron spectrum during the 
flaring period from inner blob. The narrow EED can be reconciled as a stochastic acceleration process
via Fermi II order acceleration scenario, where randomly moving Alfv\'en waves may  accelerate particles 
in turbulent medium. 

\begin{figure}[!ht]
   \centering
   \includegraphics[angle=0,width=9cm]{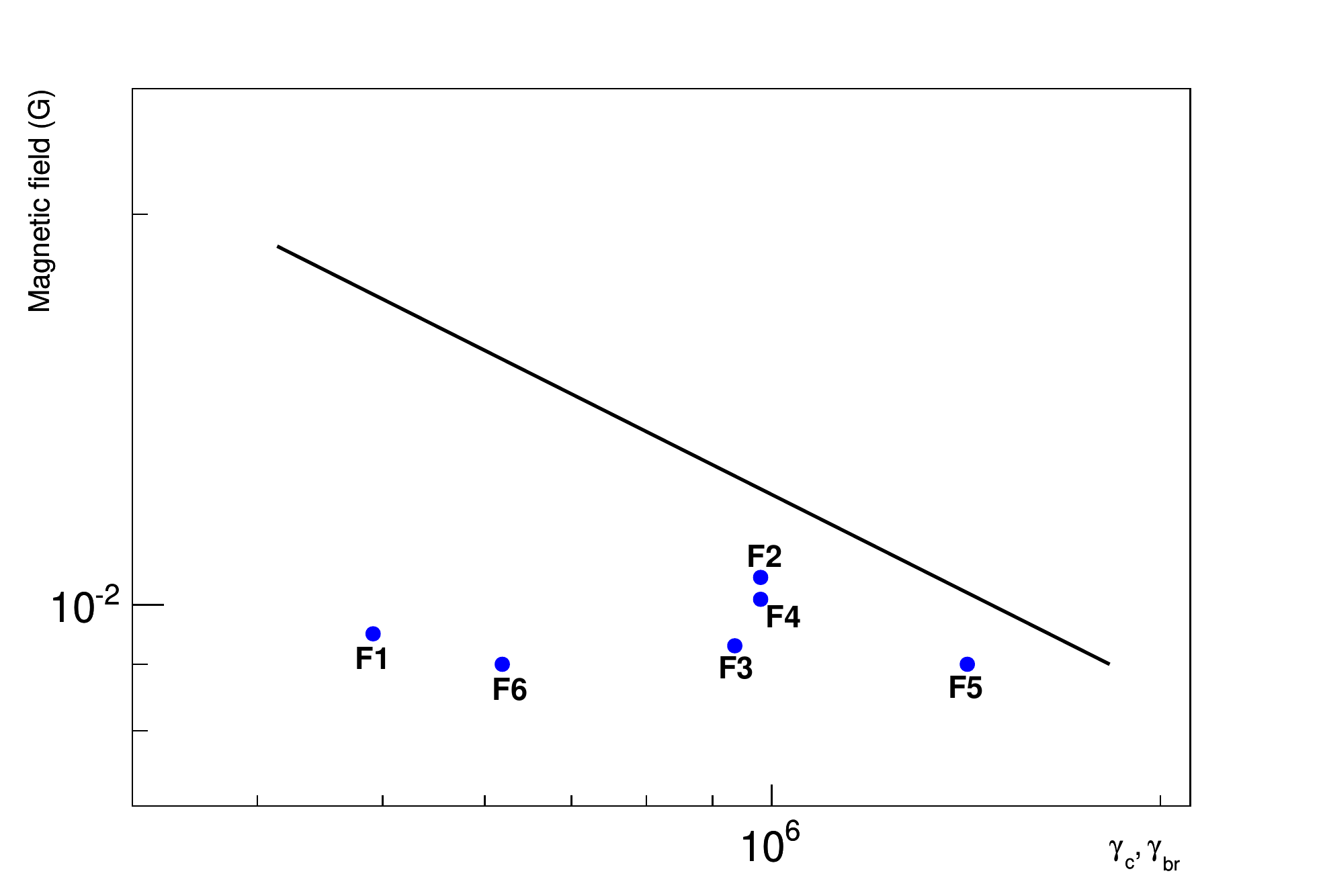}         
   \caption{The temporal evolution of model parameter B with $\gamma_{br}$ for inner zone ;
Theoretical curve shown by black continuous line is as per Eq. (9).}
   \label{FigGammaB}
\end{figure}

The observed break in the particle spectrum found by SED modeling
is at much lower energies than the expected from canonical jet model (Eq. 9) as seen from Fig.~\ref{FigGammaB}.
This observed break in particle spectrum might be an outcome of effective inverse
Compton cooling and this break appears at Lorentz factor where inverse Compton cooling 
equals adiabatic losses. In our SED modeling inverse Compton losses dominate 
synchrotron losses. Similar behavour is observed in Mkn 501 \citep{acciari2011}.
The magnetic field and electron energy density in inner region was 
found to be an order of magnitude higher than the outer region. 
Moreover, electron spectra were found much softer than inner blob. The particle spectra in outer 
blob might be outcome of shock acceleration where particles are cooled through synchrotron and IC losses.

With increase in the flux synchrotron peak is found to shift to right and vice-versa i.e. 
 'bluer when brighter' behaviour of the source during this period. Similar 
behaviour is reported by \citet{HG}. Among the six states F2, F3 and F4 states 
sampled the peak of the flare. During F3 and F4 states the high energy $\gamma$-ray
flux is comparable while there is a significant increase in the X-ray flux as the state 
evolved from F3 to F4. It was then followed by the decrease in the flux in both 
synchrotron and IC region as the flare decays in F5 and F6 state.

We are able to reproduce SEDs satisfactorily with broken powerlaw electron distribution 
with ($p_{2}$ - $p_{1}$) $\sim$ 1, for both inner and outer regions. This corresponds 
to spectral index change of $\Delta$ $\alpha$ = 0.5 for synchrotron emission, which is expected in 
the case of canonical cooling break in homogeneous model. We could reproduce all the flaring states 
and one quiescent state with Doppler factor of $\sim$ 15 for both comoving plasma 
blobs. Higher $\gamma_{min}$ seen in inner blob compared to the outer blob suggests 
the narrower EED of inner blob. The narrow EED is responsible for producing the hard 
spectrum \citep{tava, kata, lefa}. During the flaring activity inner 
blob is mostly responsible for synchrotron emission for all the states while outer blob 
is responsible for optical/UV and low energy $\gamma$-ray (0.1 - $\sim$ 3 GeV).
Except for F2 state, the outer blob contributes to observed SEDs significantly for 
all the states including quiescent one in reproducing both the humps. However in F2 state, 
outer blob was required to reproduce the optical/UV emission from source. We see clear 
plateau in this low energy $\gamma$-ray in F1, F3 and F4 states. Such plateau will be 
seen if the observed emission is sum of the emission from different region \citep{amit}.

\begin{acknowledgements}
In this paper, we used Enrico, a community-developed Python package to simplify 
Fermi-LAT analysis \citep{enrico}. This research has made use of the XRT Data 
Analysis Software (XRTDAS) developed under the responsibility of the ASI Science 
Data Center (ASDC), Italy. Data from the Steward Observatory spectropolarimetric 
monitoring project were used. This program is supported by Fermi Guest Investigator 
grants NNX08AW56G, NNX09AU10G, NNX12AO93G, and NNX15AU81G. In this research we 
used data from the OVRO 40-m monitoring program \citep{ovro} which is supported 
in part by NASA grants NNX08AW31G, NNX11A043G, and NNX14AQ89G and NSF grants 
AST-0808050 and AST-1109911.
We thank A. R. Rao for his useful suggestions and comments.
\end{acknowledgements}

\bibliographystyle{aa}
\bibliography{1ES1959+650}

\end{document}